\documentclass[preprint2]{aastex}
\usepackage{picture}

\newcommand{\gtabouteq}{\,\hbox{\raise 0.5 ex \hbox{$>$}\kern-.77em 
                    \lower 0.5 ex \hbox{$\sim$}$\,$}}       
\newcommand{\ltabouteq}{\,\hbox{\raise 0.5 ex \hbox{$<$}\kern-.77em 
                     \lower 0.5 ex \hbox{$\sim$}$\,$}}
\newcommand{\hb}{\hfill\break}
\newcommand{\rasec}
 {\hbox{$\,$\raise 0.6 ex \hbox{\rm s}\kern-.35em
                  \lower 0.0 ex \hbox{.}$\,$}} 
\newcommand{\decsec}
{\hbox{$\,$\raise 0.0 ex \hbox{$^{\prime\prime}$}\kern-.45em
                  \lower 0.0 ex \hbox{.}$\,$}}

\slugcomment{To appear in the Astronomical Journal}

\shorttitle{CHANG-ES: I}
\shortauthors{Irwin et al.}

\begin{document}


\title{Continuum Halos in Nearby Galaxies -- an EVLA Survey \\
    (CHANG-ES) -- II: First Results on NGC~4631}


\author{Judith Irwin\altaffilmark{1}, Rainer Beck\altaffilmark{2},
R. A. Benjamin\altaffilmark{3}, Ralf-J{\"u}rgen Dettmar\altaffilmark{4},
Jayanne English\altaffilmark{5},\\
George Heald\altaffilmark{6}, Richard N. Henriksen\altaffilmark{7},
Megan Johnson\altaffilmark{8},
Marita Krause\altaffilmark{9}, Jiang-Tao Li\altaffilmark{10},
Arpad Miskolczi\altaffilmark{11},
Silvia Carolina Mora\altaffilmark{12},
E. J. Murphy\altaffilmark{13}, Tom Oosterloo\altaffilmark{14},\\
Troy A. Porter\altaffilmark{15}, Richard J. Rand\altaffilmark{16},
D. J. Saikia\altaffilmark{17},
Philip Schmidt\altaffilmark{18},\\
A. W. Strong\altaffilmark{19}, Rene Walterbos\altaffilmark{20},
Q. Daniel Wang\altaffilmark{21}}
\and
\author{Theresa Wiegert\altaffilmark{22}}


\altaffiltext{1}{Dept. of Physics, Engineering Physics \& Astronomy, 
Queen's University, Kingston, ON, Canada, K7L 3N6, {\tt irwin@astro.queensu.ca}.}
\altaffiltext{2}{Max-Planck-Institut f{\"u}r Radioastronomie, Auf dem H{\"u}gel 69,
53121, Bonn, Germany,
{\tt rbeck@mpifr-bonn.mpg.de}.}
\altaffiltext{3}{Dept. of Physics, University of Wisconsin at Whitewater, 800 West
Main St., Whitewater, WI, USA, 53190,
{\tt benjamin@wisp.physics.wisc.edu}.}
\altaffiltext{4}{Astronomisches Institut, Ruhr-Universit{\"a}t Bochum, 44780 Bochum,
 Germany,
{\tt dettmar@astro.rub.de}.}
\altaffiltext{5}{Department of Physics and Astronomy, 
University of Manitoba, Winnipeg, Manitoba, Canada, R3T 2N2,
{\tt jayanne\_english@umanitoba.ca}.}
\altaffiltext{6}{Netherlands Institute for Radio Astronomy (ASTRON), 
Postbus 2, 7990 AA, Dwingeloo, The Netherlands,
{\tt heald@astron.nl}.}
\altaffiltext{7}{Dept. of Physics, Engineering Physics \& Astronomy, 
Queen's University, Kingston, ON, Canada, K7L 3N6, {\tt henriksn@astro.queensu.ca}.}
\altaffiltext{8}{National Radio Astronomy Observatory, P. O. Box 2, Greenbank, WV, USA, 24944,
{\tt mjohnson@nrao.edu}.}
\altaffiltext{9}{Max-Planck-Institut f{\"u}r Radioastronomie,  Auf dem H{\"u}gel 69,
53121, Bonn, Germany,
{\tt mkrause@mpifr-bonn.mpg.de}.} 
\altaffiltext{10}{Dept. of Astronomy, University of Massachusetts, 710 North
Pleasant St., Amherst, MA, 01003, USA, 
{\tt jiangtao@astro.umass.edu}.} 
\altaffiltext{11}{Astronomisches Institut, Ruhr-Universit{\"a}t Bochum, 44780 Bochum, Germany, 
{\tt miskolczi@astro.rub.de}.}
\altaffiltext{12}{Max-Planck-Institut f{\"u}r Radioastronomie,  Auf dem H{\"u}gel 69,
53121, Bonn, Germany,
{\tt cmora@mpifr-bonn.mpg.de}.} 
\altaffiltext{13}{Observatories of the Carnegie 
Institution for Science, 813 Santa Barbara Street, Pasadena, CA, 91101,
USA,  {\tt 
emurphy@obs.carnegiescience.edu}.} 
\altaffiltext{14}{Netherlands Institute for Radio Astronomy (ASTRON), Postbus 2,
7990 AA, Dwingeloo, 
The Netherlands, {\tt oosterloo@astron.nl}.} 
\altaffiltext{15}{Hansen Experimental Physics Laboratory, Stanford University, 
452 Lomita Mall, Stanford, CA, 94305, USA, {\tt tporter@stanford.edu}.}
\altaffiltext{16}{Dept. of Physics and Astronomy, University of New Mexico, 
800 Yale Boulevard, NE, Albuquerque, NM, 87131, USA, {\tt rjr@phys.unm.edu}.} 
\altaffiltext{17}{National Centre for Radio Astrophysics, 
TIFR, Pune University Campus, Post Bag 3, Pune, 411 007, India,
 {\tt djs@ncra.tifr.res.in}.}
\altaffiltext{18}{Max-Planck-Institut f{\"u}r Radioastronomie,  Auf dem H{\"u}gel 69,
53121, Bonn, Germany,
{\tt pschmidt@mpifr-bonn.mpg.de}.}
\altaffiltext{19}{Max-Planck-Institut f{\"u}r extraterrestrische Physik, 
Garching bei M{\"u}nchen, Germany, {\tt aws@mpe.mpg.de}.}
\altaffiltext{20}{Dept. of Astronomy, New Mexico State University, 
PO Box 30001, MSC 4500, Las Cruces, NM 88003, USA, {\tt rwalterb@nmsu.edu}.}
\altaffiltext{21}{Dept. of Astronomy, University of Massachusetts, 710 North
Pleasant St., Amherst, MA, 01003, USA, 
{\tt wqd@astro.umass.edu}.}
\altaffiltext{22}{Dept. of Physics, Engineering Physics \& Astronomy,
Queen's University, Kingston, ON, Canada, K7L 2T3, 
{\tt twiegert@astro.queensu.ca}.}


\begin{abstract}
We present the first results from the CHANG-ES survey, a new survey of 35 edge-on
galaxies to search for both in-disk as well as extra-planar radio continuum emission.
CHANG-ES is exploiting the new wide-band, multi-channel
capabilities of the Karl G. Jansky Very Large Array (i.e. the Expanded Very Large Array, or EVLA)
with observations in two bands centered at 1.5 and 6 GHz
in a variety of array configurations with full polarization.
 The motivation and science case for
the survey are presented in a companion paper (Paper I).  These first results are based on
C-array test observations in both observing bands of the well-known radio
halo galaxy, NGC~4631.  In this paper, we outline the observations and the data reduction steps
that are required for wide-band calibration and mapping of EVLA data, including polarization.

With modest on-source observing times (30 minutes
at 1.5 GHz and 75 minutes at 6 GHz
for the test data) we have
achieved best rms noise levels of 22 and 3.5 $\mu$Jy beam$^{-1}$
at 1.5 GHz and 6 GHz, respectively.  New disk-halo features have been detected, among them two at 1.5
GHz that appear as loops in projection. 

 We present the first 1.5 GHz
spectral index map of NGC~4631 to be formed from a single wide-band observation
in a single array configuration.
This map represents tangent slopes to the intensities within the band centered at 
 1.5 GHz, rather than fits across widely separated
frequencies as has been done in the past and is also the highest spatial resolution spectral index
map yet presented for this galaxy. The average spectral index in the disk is
 $\bar\alpha_{1.5 GHz}\,=\,-0.84\,\pm\,0.05$ indicating that the emission is largely non-thermal,
but a small global thermal contribution
is sufficient to explain a positive curvature term in the
spectral index over the band.  Two specific star forming regions have spectral indices that are consistent
with thermal emission. 

Polarization results (uncorrected for internal Faraday rotation) are consistent with previous
observations and also reveal some new features.  On broad
scales, we find strong
 support for the notion that magnetic fields constrain 
the X-ray emitting hot gas.
\end{abstract}





\section{Introduction}
\label{sec:introduction}

We present results of radio continuum observations of the edge-on galaxy,
NGC~4631, using the
Karl G. Jansky Very Large Array (hereafter, the Expanded Very Large Array, or 
EVLA) in its C configuration.  These are the first results from a new
survey, Continuum Halos in Nearby Galaxies -- an EVLA Survey
(CHANG-ES), whose motivation, science goals, galaxy sample criteria, and expectations
for improvements over previous surveys are described in a companion paper
\citep[][hereafter, Paper I]{irw12a}.  The sample consists of 
35 nearby edge-on galaxies and observations are being carried out in 
two frequency bands (1.5 GHz and 6 GHz, i.e. in L-band and C-band, respectively) 
in all polarization products.  The survey includes observations in three EVLA array
configurations (B, C, and D) but the test data presented here were obtained at
both frequencies in the C configuration only. 

 In this second paper, our goals are to outline the observations and data reduction
procedures used for the test data (Sect.~\ref{sec:n4631})
and to present the initial scientific results for
NGC~4631 (Sect.~\ref{sec:results}).  
We pay particular attention to the data reduction
procedures, especially with respect to differences that are introduced by using the
wide EVLA
frequency bands that are now available. 

The galaxy, NGC~4631, was chosen 
 because of its extensive radio continuum halo
which has been known for some time \citep{eke77}.  For a recent summary of
previous observations of this galaxy, see \citet{irw11}.
In this paper, we refer to
the region, $0.2\,\ltabouteq\,z\,\ltabouteq\,1$ kpc, as the
disk-halo interface and use `halo' for emission on
larger scales ($z\,\gtabouteq\,1$ kpc). 
`High-latitude' or `extraplanar' are also used to describe either
of these components.  It is worth keeping in mind that
halos are not necessarily smooth since substructure is generally
observed, depending on the spatial scales that are probed.

\section{EVLA Observations of NGC~4631}
\label{sec:n4631}

\subsection{Observations}
\label{subsec:n4631-obs}

Observations of NGC~4631 were carried out on 11 Nov. 2010 with the pointing center
set to the center of the galaxy at
RA = 12$^{\rm h}$ 42$^{\rm m}$ 08$\rasec$01, DEC = 32$^\circ$ 32$^\prime$ 29$\decsec$4
 The observing
set-up is summarized in Table~\ref{table:obs-n4631}.  
All EVLA observations (including continuum)
 are now done in spectral line mode which facilitates the excision of
 radio frequency interference (RFI) from the data,
effectively enables broader uv coverage via multi-frequency synthesis techniques
(Sect.~\ref{subsec:imaging}) and permits the extraction of spectral index information
(Sect.~\ref{sec:spectral_index})\footnote{It will also allow an analysis of
Faraday rotation, to be described in future papers; see Paper I for related science.}. 
 
At the time of the observations,
the WIDAR correlator was restricted to a maximum of 8 spectral windows (spws) per 
base-band, i.e. per AC or BD intermediate
frequency (IF).  Each spectral window has its own
bandpass response, and contains 64 channels.
With AC and BD IFs contiguous in frequency, the result was 1024 spectral channels centered at 1.5 GHz
(L-band)
for a total bandpass of 512 MHz,
and 1024 channels centered at 6.0 GHz (C-band) for a total bandpass of 2.048 GHz.

Clearly, at the lower frequency, the achievable bandwidth cannot be higher than the frequency
itself, but the L-band
bandwidth (more so than C-band) also
 suffers from RFI\footnote{A list of known RFI is given at
{\tt https://science.nrao.edu/facilities/evla/observing/rfi}.}, restricting
the bandwidth further.  In fact, although we avoided known RFI at the ends of the band,
we found that the L-band RFI, especially
in the 1.52 to 1.64 GHz region within our band, to be persistent 
 throughout our 
observations.  Together with other flagging, this reduced the effective
L-band bandwidth to approximately 300 MHz and has prompted a modification of the 
correlator set-up for subsequent observations. 
The strength of the RFI in L-band achieved a maximum value which was 300 times that of
3C286 (the flux calibrator) so that even the sidelobes of the RFI swamped that of our
astronomical sources at some frequencies.  
Experimentation with flagged and unflagged data showed that
it was not, in general, possible to salvage the long-spacing uv data while flagging only
the short-spacings; once RFI was detected, all uv data had to be removed over the time
and frequency ranges that were affected.  RFI at C-band was also present, but much less
severe than in L-band.

Observations were carried out in the standard fashion, including scans of a flux
calibrator (which was also used as a bandpass calibrator as well as for
the determination of absolute polarization position angle), 
a `zero-polarization' calibrator (to determine the instrumental polarization)
and a phase calibrator.  
The flux calibrator  
was observed in 3 independent scans at
each frequency for these test observations. Only a single scan is necessary, in principle,
but these data were intentionally `over-calibrated' for consistency checks. 
The zero-polarization calibrator was observed once at each frequency. 
 
The L-band observation of NGC~4631 was carried out in a single time block containing
one on-source scan
flanked by phase calibrator scans (plus the flux and zero-polarization calibrator).  
The C-band observation of NGC~4631 with its calibrators was similarly observed in 
a single time block, but separated into
3 individual scans interspersed with phase calibrator scans.  Observations at
each frequency were carried out within single blocks of total observing time in
order to minimize time spent on overheads.  For the regular data, however,
attempts are being made, where possible, to split the observing time for given galaxy at each 
frequency into two observing blocks that are widely separated in hour angle; more time is then spent on
overheads but the result
should be improved uv coverage with a better-behaved beam, as well as lower off-axis 
instrumental polarization (Sect.~\ref{subsec:pol-cal}).

\subsection{Total Intensity Calibration}
\label{subsec:n4631-calibration} 
The data were reduced using the Common Astronomy Software Applications (CASA) 
package\footnote{Available at {\tt http://casa.nrao.edu}.},
unless otherwise indicated. As improvements to CASA are
frequently being made during EVLA commissioning, a number of versions of CASA were
employed in the data reductions, in particular versions that were extant over the
period May 2011 to July 2011.  The following steps were carried out for both the
L-band and C-band data separately.

On-line Hanning smoothing is no longer an option at the EVLA, so all data were
first Hanning smoothed to minimize Gibbs ringing in frequency, a step which significantly
ameliorated RFI detection throughout the band. As a result, the frequency resolution
is twice the channel width, the latter listed in Table~\ref{table:obs-n4631}\footnote{If corrections are
required for antenna position or antenna-based delays, they would be carried out prior to the
Hanning smoothing step.}.
A significant amount of time was then spent flagging RFI.  Experimentation with
automatic
flagging routines 
had not yet provided results that were as dependable as manual flagging, requiring that
flagging be done by hand; however, such routines are improving and should shorten the data reduction
process in the future.

The flux density, I$_\nu$, (Stokes I) of the flux calibrator,
3C~286, was first set for the frequency, $\nu$, of each spectral window using a known model
and the `Perley-Taylor-99' 
calibration scale\footnote{See {\tt http://www.vla.nrao.edu/astro/calib/manual/baars.html}.}.
Since each spectral window 
of each antenna has its own bandpass response, bandpass calibrations were required, again
using 3C~286. However, since the data are vector-averaged when producing an average
bandpass for each spectral window, it is necessary to first carry out an initial gain and phase 
calibration with time; applying the solution ensures that phase variations that may occur
over the different scans of
3C~286 when it is observed through different elevations do not result in decorrelation when
vector-averaging occurs.  With the initial gain/phase solutions applied, 
bandpasses were then determined, one for
each spectral window of each antenna, and
the initial gain and phase solution is no longer used.  

The gain and phase were then calibrated with time for each of the flux, zero-polarization, and
phase calibrators, where an antenna-based solution is found for each spectral window and
each separate scan, followed by bootstrapping of the flux from the flux calibrator to
both the zero-polarization (OQ~208) and phase (J1221+2813) calibrators for each spectral window.
At L-band, the flux densities of OQ~208 and  J1221+2813 varied systematically over the band 
from 0.72 to 1.16 Jy (positive spectral index\footnote{$S_\nu\,\propto\,\nu^\alpha$.}
 with an error of about 0.3\% on each spectral window value)
and from 0.55 to 0.53 Jy (negative spectral index with an error of about 0.7\% on each spectral window value),
respectively.
At C-band, the flux densities of the two calibrators (in the same order)
varied 
from 2.4 to 1.9 Jy (negative spectral index with an error of about 0.4\% on each spectral window value)
and from
 0.46 to 0.44 Jy (negative spectral index with an error of about  0.5\% on each spectral window value),
respectively.
The changing spectral index of 
OQ~208 (positive in L-band and negative in C-band) is in 
good agreement with the known spectrum which peaks around
4 GHz \citep[][or the Nasa Extragalactic Database]{cen06}.

The bandpass and gain/phase solutions as a function of time
 were then applied to all sources, antenna by antenna,
and spectral window by spectral window, with the gain/phase solutions from the phase calibrator interpolated
in time when applied to NGC~4631\footnote{In practice, application of the calibration tables revealed
that more flagging for RFI was required, necessitating a second iteration through the
calibration.  The fluxes quoted in this section apply to the final iteration.  At various
stages in the data reduction, occasionally additional flagging of RFI was carried out.  
Wherever such flagging
affected the calibration or imaging, the necessary steps were redone.}.

\subsection{Polarization Calibration}
\label{subsec:pol-cal}


Polarized intensity, P, and the polarization angle on the sky,
$\chi$, require calibration of Stokes Q and 
U.  To initially set Stokes Q and U for 3C~286,
the absolute position angle on the sky is assumed to be constant at 
 $\chi\,=\,33$ deg, but the percentage polarization 
varies with frequency.  We obtained detailed EVLA data 
on this calibrator (R. Perley, private communication) and fit a curve 
through the L-band data
of the form,  
\begin{equation}
P_\nu = -1.443\,\times\,10^{-6}\,\nu^2 + 0.006026\,\nu + 3.858
\end{equation}
 where 
$P_\nu$ is the percent polarization 
 at frequency, $\nu$ (MHz).  
In C-band, only a linear fit was necessary,
\begin{equation}
P_\nu = 0.0001583\,\nu + 10.57
\end{equation}
($\nu$ in MHz).
The estimated error, including measurement errors and fitted curves is of order 0.3\%.
Q and U were then calculated at the frequency of each spectral window via
\begin{eqnarray}
{\rm Q}_\nu &=& (P_\nu/100)*{\rm I}_\nu*cos(2\chi)\\
{\rm U}_\nu &=& (P_\nu/100)*{\rm I}_\nu*sin(2\chi)
\end{eqnarray}
where the total intensity, I$_\nu$, has previously been set via a
known model (Sect.~\ref{subsec:n4631-calibration}).

There is leakage between the right circularly polarized (R) and left circularly
polarized (L) feeds at the EVLA and this leakage must be determined in order to
calibrate the cross-terms (RL, LR) from which Q and U are calculated\footnote{Q =
(RL + LR)/2, U = (RL - LR)/2i.}.  The
leakage terms (referred to as the `D' terms) were determined from the
`zero-polarization' source, 
OQ~208\footnote{{\tt http://evlaguides.nrao.edu/index.php?title=Category:Polarimetry}.}
after first applying the previously determined gain/phase and bandpass solutions.

Finally, the absolute polarization angle on the sky (the R-L phase difference)
 was calibrated from the known position angle of 
 3C~286, after first applying  solutions for
 gain/phase,  bandpass, and D terms. 

All corrections were then applied to the
calibrators and the source (gain/phase, bandpass, D terms, and 
absolute position angle corrections).

The polarization calibration described in this section corrects for ionospheric Faraday
rotation, provided that this rotation is constant with time and position in the sky.
As neither may be the case, it is important to determine the magnitude of such variations.
At present, there is no CASA routine to do this, but there is (the task, TECOR) in the
Astronomical Image Processing System (AIPS).  We therefore copied a small amount of 
L-band data from
each calibrator, ensuring that all observing times were represented in the copied data set,
into AIPS for this purpose.
(Corrections for the source are assumed to be the same as those for the phase calibrator, given
their close proximity in the sky.)
TECOR derives corrections for ionospheric Faraday rotation
from maps of total electron content obtained from NASA's crustal dynamics data interchange
system (CCDIS\footnote{{\tt http://cddis.gsfc.nasa.gov/}}) which
cover the time range of the observations.  These measurements provide
a crude correction since the spatial resolution is typically coarser than the source size.
Therefore, if significant fluctuations exist on smaller scales, we cannot correct for them.

In L-band, we find that the differential ionospheric Faraday rotation ranges from
a minimum of 2.365 to a maximum of 
2.805 rad m$^{-2}$ over all sources, positions, and times. This result corresponds
to a variation of 0.44 rad m$^{-2}$, or 0.74 deg at the high frequency
end of the band and 1.45 deg at the low frequency end.  Thus a typical error introduced
by ignoring time variable or position variable ionospheric Faraday rotation 
is about 1 deg and we therefore have made no additional 
correction for Faraday rotation at L-band beyond the non-differential term that has already been
 taken into account in the normal data reduction.
Moreover, since Faraday rotation $\propto\,\lambda^2$ and since the L-band and
C-band observations were close together in time,  no additional correction was necessary
in C-band either.

Finally, the accuracy of wide field linear polarization imaging is limited by
variations in instrumental polarization as a function of angle from the pointing center; that is, the 
corrections that have now been applied are accurate only for the
on-axis position.  The off-axis instrumental polarization for a `snapshot' 
observation\footnote{We take this to mean any observation
short enough that Earth rotation does not fill in uv tracks.\label{footnote:snapshot}} is estimated to
be less than a few percent within the half-power point of the primary beam\footnote{See the {\it Observational Status
Summary} at {\tt http://evlaguides.nrao.edu} or earlier references such as 
{\tt http://www.vla.nrao.edu/astro/guides/vlas/current/node35.html} and \citet{bri03}.} and
increases at larger angles.  In this paper,
our polarization results (Sect.~\ref{sec:pol_maps}) are shown only over regions that are within the primary
beam full width at half maximum (FWHM).
  In addition, when sources are tracked, the source parallactic angle 
rotates through the primary beam, minimizing the off-axis polarization errors.
Hence for these data, this effect is negligible.  Nevertheless, for future wide-field polarization mapping,
a correction will likely be required; a CASA algorithm is currently being developed to aid in this
process.

\subsection{Imaging}
\label{subsec:imaging}
Our goal is to form single I, Q, and U maps of NGC~4631 from 
each of the L-band and C-band data;
 maps of linear polarization intensity, P, and magnetic field position
angle, $\chi$, can then be formed from Q and U.  
As has always been done, the uv data are
first Fourier Transformed (FTed) into the map plane which results in an
image (the `dirty image') which is the convolution of the FT of the antenna distribution
in the uv plane (the `dirty beam') with
the source sky brightness distribution.  
The dirty beam is then deconvolved from the dirty image (i.e. it is `cleaned')
to find an image of the source.  The Clark method 
\citep{cla80} of separating
the cleaning process into major and minor cycles, treating each Stokes plane separately,
was used.

The broad bandwidths of the EVLA WIDAR correlator, however, 
introduce some new imaging
requirements and considerations that were, for the most part, unnecessary 
with the VLA. 
 These are described briefly here:\hb
{\it a)} {\it Bandwidth smearing:}  If the observations had not been in spectral line mode,
then the broad frequency band would have introduced severe chromatic aberration into
the image; this aberration worsens with distance from the map center.  However,
since spectral line mode is being universally employed at the EVLA,
bandwidth smearing is instead restricted to the spectral resolution of the data (by using
the multifrequency synthesis technique, see
next section).  For example,
for our data 
at the center of either band and at a distance from the map center that is twice the
FWHM of the primary beam, the loss in
amplitude due to bandwidth smearing is only about 1.5\%, or essentially negligible where most of
the emission occurs.\hb
{\it b)} {\it Multi-frequency synthesis (mfs):} The positions of the antennas in the uv plane
are measured in wavelengths and those positions will vary as the
frequency changes across a given band.  Consequently, if each spectral channel is gridded
separately, there is an improvement in uv coverage over a case in which channels are averaged
together.  The clean algorithm employed in CASA utilizes this multi-frequency synthesis
technique \citep[][also called `bandwidth synthesis']{con90}.\hb
{\it c)} {\it Widefield imaging:} The sensitivity of the array to emission that is
far from the map center and even  
outside of the telescope's primary beam means that the array itself is non-coplanar to
such emission.  Thus, the `w term' in the FT (the coordinate in the line-of-sight direction
towards the source), which can be ignored near the map center,
 becomes important.  CASA clean employs the `w-projection algorithm' \citep{cor08b} to
correct for this effect.\hb
{\it d)} {\it Spectral index fitting:} With wide bands, the spectral index of the source
itself, which is spatially variable, needs to be fitted during imaging 
\citep{sau94}.  In CASA's clean algorithm, this is accomplished 
by assuming that the emission at any position
can be described by, 
\begin{equation}
\label{eqn:spectralfit}
{\rm I}_\nu\,=\,
{\rm I}_{\nu_0}\,\left(\frac{\nu}{\nu_0}\right)^{\alpha\,+\,\beta\, log\left(\frac{\nu}{\nu_0}\right)}
\end{equation}
where $\nu_0$ is a reference frequency within the band (here taken to be the band center), 
I$_{\nu_0}$ is the specific intensity at the
reference frequency, $\alpha$ is the spectral index and $\beta$ describes the spectral 
curvature.
The function is expanded in a Taylor series about $\nu_0$, resulting in a polynomial fit to
the spectrum.  
The fit includes all unflagged channels across any given band and is not done on a `per spectral window' 
basis.
In addition, the fitting is done using all data 
so that the resulting parameters correspond to a single spatial resolution (for example, the
$\alpha$ map is not formed such that the spatial resolution would vary across the
band)\footnote{At the time of writing, this is the only CASA-based fitting function that is available.}.
\citet{rau11} outline the 
implementation of spectral fitting and provide examples as to how the image can improve with
increasing numbers of Taylor terms, $n_T$. 
\hb
 {\it e)} {\it Multi-scale clean:} Improvements have been made over the traditional 
clean which modeled all emission as a collection of point sources.  The clean implementation
in CASA now 
allows for a multi-scale clean \citep{cor08a} which assumes that the emission can be modeled
as a collection of components over a variety of
spatial scales.  These scales are typically chosen to span the range from
an angular size of `zero' (corresponding to a point source, or the classic clean analogy) to a scale of order the
maximum spatial scale expected to be present in the emission.  Since the latter is
not always known in advance, in practice, the maximum scale size is increased
systematically until no more flux is actually cleaned out in that largest scale\footnote{If the flux
becomes negative in the largest scale, the trial is aborted and the maximum scale is decreased until
positive flux is cleaned in all scales.}. 
 For details of the CASA implementation of the
combined multi-scale, multi-frequency synthesis (ms-mfs)
clean, see \citet{rau11}.  \hb
{\it f)} {\it Primary beam:} The primary beam (PB) varies from one end of the band to the other
and, as a result, nowhere in the field of view is there a `null' as there would be for
a monochromatic primary beam \citep[][]{bha11}.  Consequently, it is the frequency-averaged PB (PB$_{avg}$)
that the final cleaned image must be corrected for.  Moreover, since the frequency dependence
of the PB imposes its own spectral index across the field of view, all spectral index maps
(Sect.~\ref{sec:spectral_index})
must be corrected for this effect as well. 
 
\smallskip

NGC~4631 was mapped taking into account 
 all of the above points. 

At both bands, a very wide field was first mapped (3 degrees for L-band, or 6 times the
primary beam FWHM; 34.1 arcmin for C-band or 4.6 times the primary beam) to 
 determine how large an image needed to be mapped to include all
sources of significant flux in the clean.  
Smaller fields were then mapped completely, as required.
For the multi-scale clean, we used typically 4 to 6 equally-spaced spatial scales, the first one
always corresponding to point sources as in the classic clean.  For example,
at L-band, scales up to 4.2 arcmin were employed, where the largest scale was chosen
according to the considerations of point {\it e)} above.
We also self-calibrated each data set once, finding that
a single amplitude + phase self-calibration resulted in a small improvement in the
rms map noise\footnote{A phase-only self-calibration did not improve the results.}.  

After extensive experimentation, we adopted 3 sets of parameters representing
3 different spatial resolutions for each frequency and display them, along
with some ancillary data, in 
Figs.~\ref{fig:L-band} and \ref{fig:C-band}; the corresponding map parameters are given in
 Table~\ref{table:map-n4631}.  

 The theoretical noise values for these observations (including
confusion estimates) are 
18 and 3.7 $\mu$Jy beam$^{-1}$ at L-band and C-band, respectively.  Our lowest rms noise
values are only 1.6 and 1.3 times higher for the total intensity images. 
There are, however, several residual linear cleaning artifacts in the L-band images
 that are visible immediately to the east and west of the galaxy's disk (see Fig.~\ref{fig:L-band}c
in particular). The conclusions that
 we present in this paper are not affected by these artifacts.
 The disconnected emission seen approximately 2 arcmin above and below the plane
in  Fig.~\ref{fig:C-band}c is likely real and part of the halo; missing low-order
spacings contribute to the well-known appearance of the galaxy's disk emission
 sitting on a negative bowl.

Note that, as is usually done,
 all displayed total intensity images are not corrected for the primary beam
in order to show images with uniform noise.
When numerical results are given, however, the results have been corrected for the
frequency-averaged primary beam (PB$_{avg}$).
Note also, that we carried out some subsequent analysis using AIPS.

\subsection{Polarization Imaging}
\label{sec:polarization}

Stokes Q and U maps were formed with the same
sets of parameters as the total intensity images, except that fewer spatial scales were 
required during cleaning, given the weaker emission in these maps.
The relevant parameters for these images are given in Table~\ref{table:map-n4631}.  In
principle, the rms noise on the Q and U maps should be the same as the I map; however,
the lower signal in the cross-hands means that residual errors associated with cleaning
the high intensity emission are no longer present.  At L-band, our lowest rms noise is
1.2 times the theoretical value and at C-band, our lowest noise matches the theoretical value.
Note that, in order to obtain sufficient signal-to-noise (S/N) in the polarized emission, 
all spectral windows were used. 


We formed polarized intensity images, 
P$=\sqrt{{\rm Q}^2\,+\,{\rm U}^2}$, for each of the uv weightings of
Figs.~\ref{fig:L-band} and \ref{fig:C-band}, correcting for the
bias introduced by the fact that P images do not obey Gaussian statistics
\citep{sim85,vai06}.
Maps of the polarization angle of the electric field vector, $\chi\,=\,(1/2)arctan({\rm U}/{\rm Q})$, were also formed
for points for which the
linear polarization was $>$ 4$\sigma$\footnote{The cutoff for $\chi$ was relaxed slightly from 5$\sigma$
in order to show the trends in vector orientation with position.}.  
When displayed, we have rotated $\chi$ by 90 degrees
to illustrate the observed (uncorrected for internal Faraday rotation)
 direction of the magnetic field.   We show the results for only a single representative
data set in the two frequency bands
(Robust = 2 for L-band and Robust = 0 + uv taper for C-band) in
Fig.~\ref{fig:pol_maps}\footnote{Note that the CASA package did not allow for variable
vector lengths, hence they are drawn as constants in these figures.}.  
Consistent with the total intensity images (Sect.~\ref{subsec:imaging})
the displayed maps are shown without primary beam corrections.

We then carried out a PB correction using the
same frequency-averaged primary beam, PB$_{avg}$, as was used for the total intensity images.
This correction is good to zeroth order, but does not take into account first order effects
which would result from 
differences between the right and left circularly polarized PBs.
First order corrections
 are not yet implemented in CASA but we expect them to be small
for this data set.  For example, if the difference
between the right and left circularly polarized PB voltage response is of order 10\%
at a distance from the map center corresponding to 30\% of the primary beam
(17.4 arcmin and 4.6 arcmin
at L-band and C-band, respectively), then the error in Q or U would be 3\% of the Q or U intensity at
that point. 

PB corrections of the linearly polarized intensity maps 
shown in  Fig.~\ref{fig:pol_maps}
of all emission greater
than 5$\sigma$
 resulted in increases in 
flux of 5\% and 40\%
 at L-band and C-band, respectively.  
Maps of $\chi$ (shown in Fig.~\ref{fig:pol_maps}) do not require PB correction.

We then formed maps of
percentage polarization for the two bands (these also do not require PB correction). 
The maps were formed for emission which was greater than
5$\sigma$ in both linear polarization intensity as well as total intensity, ensuring
that the spatial resolutions are matched. 
We do not display the L-band maps since the result had sufficient S/N  in only
20 beams and these positions were non-contiguous spatially.  
 The C-band percentage polarization
map is shown in Fig.~\ref{fig:percentpol}.  We calculated the average percentage polarization
over the region shown by measuring the total polarized flux (rather than per pixel values) and
dividing by
 the total flux, so that high values around the periphery
of the emission do not skew the result; we find an average of 7.0\%
for the result corresponding to the region displayed.  This region is mainly in the disk where
some Faraday depolarization may be occurring (Sect.~\ref{sec:pol_maps}).

\subsection{Formation of Spectral Index Maps}
\label{sec:spectral_index}
As indicated in Sect.~\ref{subsec:imaging} point {\it d)}, the wide bands that are now
available at the EVLA make it possible, in principle, to create 
spatially resolved spectral index maps, $\alpha$, 
(and curvature maps, $\beta$, if there were sufficient signal-to-noise)
over a single band with a single observation and common spatial resolution.  One could form a map of $\alpha$ for
L-band and for C-band individually, and then form such a map between the two bands, assuming a
constant spectral index between the two bands.

For observations at two frequencies with a single EVLA array, as for our C-array test observations,
however, there are some limitations.  First, as specified in
Paper I, we do not detect spatial scales at L-band and C-band that are greater
than 16.2 and 4 arcmin,
respectively\footnote{These values are for full-synthesis observations.
For short observations of $\ltabouteq$ one minute duration (snapshots) 
these values could reduce to as little as
8 and 2 arcmin, 
respectively (see {\it Observational Status Summary} at
http://evlaguides.nrao.edu).
Our observations (see
Table~\ref{table:obs-n4631}) are intermediate between a snapshot and full synthesis. Consequently,
the spatial scales quoted in the text are likely upper limits.}.
The halo of NGC~4631 at both bands has previously been observed to extend
to approximately 12 arcmin (see, for example,
Fig. 9 of \citet{hum91} and Fig. 3 of \citet{gol94}).
 It is therefore clear that not all flux has been
detected at C-band and, as Fig.~\ref{fig:L-band} illustrates, the broadest scale halo emission
has not been detected at L-band either. Future D-array observations, with additional
shorter spacings, should remedy this situation at L-band.

Using previously published total flux estimates, we can normally quantify the amount of flux that has been
missed in our observations.  However, at C-band, a wide range of
 4.85 GHz fluxes have been reported,
likely reflecting similar problems with spatial scales,
for example,  $S_{4.8GHz}\,=\,438$ mJy,   
243 mJy (see NED) and 480 mJy \citep{gol94}. 
Our primary-beam-corrected flux in C-band is $S_{6GHz}\,=\,180$ mJy.
 Using the
 average global spectral index of \citet{hum90}, $\alpha_{avg}\,=\,-0.73$,
 our frequency-adjusted
flux would be $S_{4.8GHz}\,=\,210$ mJy, indicating that there is significant flux missing
from the C-band maps.  We thus do not show our C-band spectral index maps 
in this paper.

At L-band, we do form spectral index maps.  The primary-beam-corrected
flux from these C-array observations is $S_{1.5GHz}\,=\,935$ mJy, or 77\% of the VLA D-array VLA fluxes
of 1.22 Jy at $\nu\,=\,1.49$ GHz
\citep{hum90} or
1.29 Jy from the Westerbork Synthesis Radio Telescope at 
$\nu\,=\,1.365$ GHz \citep{bra07}\footnote{The NRAO VLA Sky Survey (NVSS)
flux of 981.6 mJy has been underestimated for this galaxy \citep[see][]{con02}.}.
Since the missing flux is on the largest scales, we form spectral index
 maps at L-band beginning with the highest resolution image (Fig.~\ref{fig:L-band})a,
and then consider how this and other errors affect the result.
 Since the L-band 500 MHz 
bandwidth is much narrower than the frequency spacing which has been used in the past
to calculate $\alpha$ \citep[e.g. 610 MHz to 1412 MHz, or 4.75 GHz to 10.7 GHz, see][]{wer88},  
the values of $\alpha$ presented here
can be thought of as
tangents to the frequency spectrum at each spatial position in L-band.

Spectral index and curvature maps were formed from 
ratios of  
the Taylor expansion coefficient maps which describe
Eqn.~\ref{eqn:spectralfit}\footnote{$\alpha\,=\,TT1/TT0$ and $\beta\,=\,TT2/TT0\,-\,\alpha(\alpha-1)/2$, where $TT0$, $TT1$, and $TT2$ are maps of the first 3 Taylor terms.},
after first correcting these coefficient maps for PB$_{avg}$
and then introducing a 5$\sigma$ total intensity cut-off.
It has been shown that
increasing the number of terms to the polynomial fit will generally improve 
the result, subject to S/N considerations \citep{rau11}. For example, 
fitting both
$\alpha$ and $\beta$ allows $\alpha$ to be well-determined at the reference frequency
even though $\beta$ may not have sufficient S/N in each synthesized beam to be 
 presented as a map.
In Fig.~\ref{fig:alpha}
we show spectral index maps corresponding to the high resolution L-band data
(Fig.~\ref{fig:L-band}a) for a straight spectral index assumption (a) and then allowing
for curvature within the band (b). 


The ``mottled'' appearance of the two maps is due in part to 
the fact that point-to-point variations are significant.
For example, after subtracting the two spectral index
maps, the rms of the result is $\sigma\,\approx\,0.4$ (omitting approximately a beam width
around the perimeter of the emission) suggesting that the per-pixel error 
from the fitting is of this order.  
There are, however, variations in the map that
exceed this uncertainty. 
In addition, there are some global trends   
in $\alpha$ which are very
similar between the two maps and the average spectral indices ($\bar\alpha$, averaged over the maps)
 are in very good
agreement, differing by only 0.01.  
Clearly,   
the error bar on the spectral index decreases
with the number of points that are averaged, i.e. with increasing S/N, as one would expect;
we will return to this point in Sect.~\ref{results:spectral_index}.  


Regarding the effect of missing broad scale flux, our measured $\alpha$ will not be affected provided
 the broad scale emission has the same
spectral index as the disk and contributes a constant intensity at each position.
It is known, however, that the halo spectral index, which we associate with the broad scale
emission, is steeper than that of the disk.  A variety of measurements have been made over
different frequency ranges, but typically, off-major-axis values fall in the range,
$\alpha\,=\,-0.9\,\to\,-2$ \citep[see discussion in][]{hum90}.  

Taking the extreme case of $\alpha\,=\,-2$, we can determine how the addition of a `plateau'
of emission of this spectral index to our data would affect our measured values of $\alpha$.
If the halo emission occupies an emission ellipse of major $\times$
minor axis extent, approximately 15 arcmin $\times$ 12 arcmin \citep{hum88},
then the emission region
of Fig.~\ref{fig:alpha} occupies approximately 20\% of this area. 
Scaling the missing flux of Fig.~\ref{fig:L-band}a to this area, assuming that
 the broad scale flux uniformly fills the emission ellipse, then the missing emission corresponds
to a plateau of 0.12 mJy beam$^{-1}$, on average,
over the disk at the central frequency.
 

Using the measured average intensity from Fig.~\ref{fig:L-band}a 
over the same region as the spectral index (taking $\beta\,=\,0$) 
the addition of such a steep spectral index plateau 
would adjust our measurement of
 $\alpha$ from -0.84 to -0.88 
($\Delta\,\alpha\,=\,0.04$). This is a worst-case scenario unless
there are missing spatial scales that are {\it smaller} than 11 arcmin (the major axis
size).

Spectral index, $\alpha$, and curvature, $\beta$, maps for L-band were also formed for
the remaining mid and low resolution maps corresponding to
Fig.~\ref{fig:L-band}b and c, respectively and are shown in Fig.~\ref{fig:alpha_lowres}.
For these resolutions, we find
$\bar\alpha\,=\,-0.82$ and $\bar\alpha\,=\,-0.87$, respectively. 
The curvature, $\bar\beta$, averaged over the emission regions 
for these two resolutions is
$1.9$ in both cases, agreeing to within $\pm\,0.1$ with the result at high resolution.
If we consider fitting errors, possible errors from missing flux, and
variations between the different uv weightings, our average spectral index is 
$\bar\alpha\,=\,-0.84\,\pm\,0.05$.

As mentioned earlier, it is worth stressing that the wide bands used here allow for considerable
flexibility in `tuning' the spatial resolution of the array by appropriate use of uv weighting.
This means that the spatial resolutions of L-band and C-band data can actually
be matched exactly, even with observations in only a single array configuration. Therefore
spectral index maps can also be made between L and C-bands from observations
in a single array configuration.  Although we formed such maps for NGC~4631, we defer the
results until all missing C-band flux is recovered through additional observations (Paper I).


%



\section{Results and Discussion}
\label{sec:results}

These new EVLA observations, even though they were obtained during a limited test run
in a single array configuration
(Table~\ref{table:obs-n4631}), have revealed a number of new features not before seen in
NGC~4631.  In this section, our intention is not to make an exhaustive list of all features, but
rather summarize the more obvious or new results and to place them in the context of
previous knowledge of this galaxy.

\subsection{Total Intensity Images and Comparison with other Wavebands}
\label{sec:results_total}

The total intensity images displayed in  Figs.~\ref{fig:L-band} and \ref{fig:C-band} show
numerous extensions away from the plane at both frequencies.  

At L-band, in particular, two
of the more prominent features appear in projection as loops or partial loops (although they
could represent shells) that have been outlined in red in 
Fig.~\ref{fig:L-band}b.  Although we have noted a few cleaning artifacts in this image to the east and
west of the major axis of the galaxy (Sect.~\ref{subsec:imaging}),
 the loops do not fall within the affected regions and they are either of sufficient signal-to-noise
and/or persistent over various uv weightings (some not displayed here but formed
during experimentation with uv weightings as described in Sect.~\ref{subsec:imaging})
that we consider them to be real features.

The diameter of the larger northern loop, which is only partial in Fig.~\ref{fig:L-band}b, is 6.3 kpc.
It is also visible in the lower resolution Fig.~\ref{fig:L-band}c where it appears to `frame' 
the companion dwarf elliptical galaxy, NGC4627 which is 2.6 arcmin to the NW of the galaxy's center.
This is likely a coincidence, since we will show below that the loop is related to the hot X-ray gas.

The diameter of the smaller more easterly loop is 2.6 kpc and this loop becomes `filled in' in
 the lower resolution Fig.~\ref{fig:L-band}c.
The eastern side of this loop and the weaker emission that extends upwards from it corresponds to
the HI `worm' observed by \citet{ran93}.  The latter feature sits above a significant HI supershell
found by those authors whose diameter (3 kpc) is approximately the same as our eastern loop.  
It is likely that these features are related and are associated with 
star forming regions that are present in the underlying disk.  The closest
prominent star formation complex is located 
 at RA $\approx$ 12$^{\rm h}$ 42$^{\rm m}$ 22$^{\rm s}$, 
DEC $\approx$ 32$^\circ$ 32$^\prime$ 42$^{\prime\prime}$; it is 
  marked with a cross in the H$\alpha$ map of Fig.~\ref{fig:C-band}c.

A search of previous radio continuum images of NGC~4631
\citep{hum88,hum90,hum91,gol94,dum95,gol99,bra07,kra09,hea09} does not reveal these loops,
although the continuum image in \citet{bra07} shows hints of the larger central loop.  Since
we see extraplanar structure on all scales in this galaxy, detecting specific features requires
appropriately weighted uv data over scales that match the feature of interest. The spatial scales
highlighted by our new data have now revealed these loops and the other arcs and filaments that we see
extending from the plane.  It is possible that some of the other broad extensions (for example, 
at RA $\approx$ 12$^{\rm h}$ 42$^{\rm m}$ 12$^{\rm s}$, 
DEC $\approx$ 32$^\circ$ 31$^\prime$ 30$^{\prime\prime}$) would resolve into loops if viewed at higher
resolution and sensitivity.

The close connection between X-ray emission and the radio halo, first pointed out 
for this galaxy by
\citet{wan95} and \citet{wan01},
finds stronger support in these observations. 
 In Fig.~\ref{fig:L-band_xray},
we show the soft X-ray emission from Fig.~\ref{fig:L-band}a,
 enhanced to show extra-planar emission,
 together with the
low resolution L-band image from  
Fig.~\ref{fig:L-band}c.  In this figure, 
we see that the X-ray emission forms a loop interior to the northern radio continuum loop, suggesting
that hot gas may be confined
by magnetic pressure.  It is interesting that there is even a smaller vertical X-ray protrusion
(RA $\approx$ 12$^{\rm h}$ 42$^{\rm m}$ 07$^{\rm s}$, DEC $\approx$ 32$^\circ$ 36$^{\prime}$) 
above the gap on the north-eastern side of the radio continuum loop.  A close relationship
has been found between radio continuum and X-ray emission in the outflow of NGC~253 as well
\citep{hee11}.

We can determine whether such confinement is feasible, assuming a simple case of
unity filling factors, by estimating the minimum
energy magnetic field strength following
\citet{bec05}. Assuming a proton to electron number density ratio of 100,
a line of sight distance equal to the width of the radio loop as shown in
Fig.~\ref{fig:L-band_xray}, an isotropic field direction, and a spectral index equal
to the global average disk value,
 we find B$_{min}\,\,\approx\,\,8$ $\mu$G.  The magnetic pressure is then 
$P_{mag}\,=\,3\,\times\,10^{-12}$ erg cm$^{-3}$.  By comparison, the thermal
pressure from the hot, X-ray-emitting gas 
is $P_{th}\,=\,6.9\,\times\,10^{-13}$ \citep{wan01}, implying that magnetic pressure is
indeed sufficient to confine the thermal gas.  

At C-band, where we have higher spatial resolution 
(Fig.~\ref{fig:C-band}), we continue to see many vertical extensions.
Again, we do not discuss every feature, but point out 
the brightest extension which is on the north of the disk at 
RA = 12$^{\rm h}$ 42$^{\rm m}$ 09$^{\rm s}$, DEC = 32$^\circ$ 33$^{\prime}$ 10$^{\prime\prime}$,
and best seen in Fig.~\ref{fig:C-band}b.  This extension
is directly below the west side of the large northern loop seen in Fig.~\ref{fig:L-band}b.  
This feature protrudes 1 kpc above the main emission below it, or about 1.6 kpc from mid-plane.
The detail and numerous extensions observed on this map reveal the complexity of the disk-halo
interface in this galaxy.

The disconnected `patches' of emission visible in Fig.~\ref{fig:C-band}c 
 above and below the plane appear to represent brighter parts of 
the broader scale halo emission which has not been detected in its entirety in
our data due to missing short spacings at C-band (Sect.~\ref{subsec:imaging}).  The feature at
RA $\approx$ 12$^{\rm h}$ 42$^{\rm m}$ 23$^{\rm s}$, 
DEC $\approx$ 32$^\circ$ 34$^{\prime}$ 30$^{\prime\prime}$
for example, corresponds to the
HI worm mentioned above.

The well-known close relationship between radio continuum, H$\,\alpha$, and CO emission
is also evident from Fig.~\ref{fig:C-band}, to the extent that `kinks', widenings and narrowings
along the major axis are correlated between these components.  Radio continuum and H$\alpha$ peaks
are nicely coincident in the wider `bulge' region \citep{irw11} about 3 arcmin to the east of
the galaxy's center within which the bright star formation complex mentioned above is located. 

\subsection{Polarization Maps}
\label{sec:pol_maps}

The linear polarization maps with observed magnetic field vectors are shown in Fig.~\ref{fig:pol_maps} and 
the percentage polarization with observed magnetic field vectors for C-band is shown in Fig.~\ref{fig:percentpol}.
Recall that the Q and U images have lower rms than the total intensity images
(Sect.~\ref{sec:polarization}) and hence some polarized emission can be seen outside of
regions of total intensity.  

All maps have been corrected for foreground
ionospheric Faraday rotation, where necessary
(Sect.~\ref{subsec:pol-cal}), but no Faraday rotation correction has been done for the
galaxy itself\footnote{We will carry out such an analysis in a future paper.}; 
therefore, especially in the disk at L-band, we expect significant Faraday depolarization,
as indicated in Sect.~\ref{sec:polarization}.
This is evident in Fig.~\ref{fig:pol_maps}a which shows little polarized intensity in the
disk at L-band.
Previous L-band images \citep[see especially][]{hea09,hum91} similarly
show little or no polarization right in the disk.  The halo, however, should not
be as strongly affected.
The effects of
 Faraday rotation should also be lower at C-band (Fig.~\ref{fig:pol_maps}b) but will not be negligible
in the disk. For example, rotation measures of up to $\pm$ 300 rad/m$^2$ have been found for NGC~4631 in a
larger 85 arcsec beam \citep{kra04}, corresponding to 43 degrees of rotation at 6 GHz; the value could be higher
still
in our data set, given our smaller beam.
 Again, away from the plane, Faraday rotation should be minor, as has been
demonstrated
previously for the halo region of the edge-on galaxy, NGC~5775 \citep[][cf. Figs. 3 and 5 of that paper]{soi11}.
Nevertheless, we prefer to be cautious about our comments regarding the magnetic field orientation in the following
discussion.

At L-band, 
the most prominent halo feature is a highly linearly polarized spur
located at RA $\approx$ 12$^{\rm h}$ 42$^{\rm m}$ 20$^{\rm s}$, 
DEC $\approx$ 32$^\circ$ 34$^\prime$ 30$^{\prime\prime}$.  This spur is visible in the linearly polarized map of 
 \citet{hea09} and also in the lower resolution 5 GHz map of \citet{gol94}.  
The spur is directly above the eastern radio continuum loop and may be related to the disturbance
that has caused that loop as well as the HI supershell below it (Sect.~\ref{sec:results_total}).
However, the galaxy also displays an V-shaped field structure 
on the northern side (or `X-shaped' globally, see above references) and this spur may be associated with
the eastern side of that larger feature.

At 6 GHz (Fig.~\ref{fig:pol_maps}b, Fig.~\ref{fig:percentpol}), it is interesting that
the observed magnetic field orientation within the disk is parallel to the disk. 
Previously published maps have shown 
the magnetic fields near the center of NGC~4631 to be perpendicular
to the disk \citep[for example, see the 8 GHz image of][at 84 arcsec resolution]{kra04}, although
at larger radii the fields are parallel to the disk again \citep{kra09}.
 Our 12 arcsec observations and omission of the largest spatial scales are targetting different scales
and may now be revealing a disk-parallel component near the galaxy's center, although Faraday-corrected
maps are still required to confirm this.  
Away from the disk, the magnetic field tends to become perpendicular to the plane, in agreement
with previous observations.  Extra-planar polarized emission both to the north-east and north-west of
the galaxy's center agrees in general with the image of \citet{gol94} at 5 GHz.  

A final point of interest is the
sharpness with which the {\it apparent} magnetic field orientation changes (90 degrees within a
single beam) at
RA $\approx$ 12$^{\rm h}$ 42$^{\rm m}$ 12$^{\rm s}$, 
DEC $\approx$ 32$^\circ$ 32$^\prime$ 30$^{\prime\prime}$ (see Fig.~\ref{fig:percentpol}).
 This change occurs at
the location of a bright HII region complex visible in
Fig.~\ref{fig:C-band}c which has been labelled CM67 by \citet{gol99}.
If such a change is still visible after having corrected for Faraday rotation, 
 then it lends support for the notion
that the magnetic field is swept outwards with the outflow from specific star forming
regions.  Note that there are radio continuum spurs above and below this location (Fig.~\ref{fig:C-band}c).
If the change is not intrinsic but is due to  
 Faraday rotation alone, then
 either the ordered magnetic field in the line of sight 
has changed sharply at this position or the electron density has changed abruptly, or both.  In either case,
CM67 is an important region for follow-up studies with applied rotation measure corrections.

\subsection{Spectral Index Images}
\label{results:spectral_index}

Our L-band spectral index maps (Figs.~\ref{fig:alpha} and \ref{fig:alpha_lowres})
are the first that have been formed from {\it within} an observing band for an edge-on
galaxy and  represent a tangent to the
spectrum at 1.5 GHz, rather than an average between broadly separated
frequencies as has been done in the past. 

The average spectral index is
$\bar \alpha$ = - 0.84 $\pm$ 0.05 (Sect.~\ref{sec:spectral_index}) and
therefore the L-band maps are clearly dominated by non-thermal emission.
This result is typical of the values found in
other galaxies \citep{nik97} and
 the Milky Way 
\citep{rei04,str11}.
Previous values for the disk of NGC~4631 cover a rather wide range \citep[see][]{hum90}, for example,
-0.68 (from 610 MHz and 10.7 GHz measurements),
-0.45 (from 327 MHz and 1.49 GHz),
-0.69 (from 327 MHz, 610 MHz, 1412 MHz, and 10.7 GHz data), and
-0.9 (from 610 MHz and 1412 MHz).

Also, although the global spectral index of NGC~4631 has been studied extensively
\citep{kle84,suk85,wer88,hum90,poh90}
the only previous
spectral index {\it maps} that have been published can be found in
\citet{wer88} whose highest resolution was 58 arcsec. Our maps are
the highest resolution spectral index maps yet presented for this galaxy.

The average, global spectral curvature is positive
 ($\beta\,=\,1.9$, Sect.~\ref{sec:spectral_index}).
The most likely explanation is the contribution of thermal emission from
the disk.  To see whether this is feasible, 
we use the thermal flux density estimate of \citet{gol99}
of $F_{T}(4.86 ~GHz) = 56$ mJy who makes use of the H$\alpha$ luminosity of NGC~4631 and adjusts 
upwards to account (roughly) for dust obscuration.  
Extrapolated to 1.5 GHz with a $\nu^{-0.1}$ frequency dependence gives 
$F_{T}(1.5 ~GHz) = 63 $ mJy with a variation of 2 mJy across L-band.
The curvature that we have found from Eqn.~\ref{eqn:spectralfit} corresponds to an increase of 2\%
at the upper frequency band edge in comparison to a non-curving spectral index.
 With a total flux density of 935 mJy at the band center (Sect.~\ref{sec:spectral_index}),
a curved spectral index results in an increase of 16 mJy at the high frequency
end of the observing band, over the value that would be observed if the spectral index
were constant.  This increase
 is
 less than the estimated thermal flux density at the upper end of the band, so it is clear that a thermal
contribution is capable of accounting for the observed curvature.  However, thermal emission
contributes, on average, less than 7\% of the flux at L-band, assuming that the thermal estimate
of \citet{gol99} can be applied.

An interesting result is the detail with which we have mapped the
spectral index distribution in a single observation in a single frequency
band. 
As indicated in Sect.~\ref{sec:spectral_index}, 
 the uncertainty at any point is approximately $\pm\,$0.4, though it becomes lower 
depending on the size of the region over which averages are taken.
For the average spectral index over all maps, the uncertainty is smaller, i.e.
 $\pm\,$0.05.
Consequently, in the following, we point out only values of $\alpha$ that are
significant in comparison to the 
uncertainties which have been appropriately averaged over the region of interest
and for which there are similar trends for all maps shown in 
 Figs.~\ref{fig:alpha} and \ref{fig:alpha_lowres}.


For example, contrast in the spectral index is observed in the
region of the eastern loop (see dashed arc in Fig.~\ref{fig:L-band}b).
The contrast is most clearly seen in Figs.~\ref{fig:alpha_lowres}a and b. 
The average spectral index along the sides of the loop itself
 (all maps) is $\bar\alpha\,=\,-1.3$ and this value 
 does not differ significantly
from the underlying disk at the loop `footprints'.
 In 
the regions interior to the loop and just exterior to it, on the other hand, the values are much steeper 
(darker regions in Fig.~\ref{fig:alpha_lowres}a and b).  The average spectral index interior
to the loop is $\bar\alpha\,=\,-2.1$.
  Both values indicate that a thermal contribution will be negligible at these locations
and, as we did in  Sect.~\ref{sec:results_total}, we estimate a minimum
energy magnetic field strength using the same assumptions as in that section
but with these measured spectral indices.
The result is B$_{min}\,\,\approx\,\,17$ $\mu$G along the loop\footnote{By comparison, a classical calculation 
\citep{pac70} using a heavy particle to
electron ratio of 40, and frequency limits from $10^7$ to $10^{11}$ Hz, results in 
B$_{min}\,\,\approx\,\,10$ $\mu$G.}.  

This is not the first time that spectral index contrasts 
or steep spectral indices have been observed in galaxies that experience outflows. 
For example, \citet{lee01} show that the spectral index of radio continuum
features that are located along HI shells in
NGC~5775 is flatter than in the interior and that this flattening cannot be
due to a thermal contribution.  As in NGC~4631, the spectral index
along the loop itself is not significantly different from that of the underlying
disk at its location.  Moreover,
although such steep spectral indices are not generally seen in our own Galaxy 
\citep{rei04, str11,gho11},
\citet{hee11} also find regions of very steep spectral index 
in the galaxy, NGC~253, i.e. $\alpha\,=\,$$-2.0 \,\pm\, 0.2$ is seen within the outflow cone 
originating from the nuclear region of that
galaxy.  
NGC~4631 is known to be experiencing widely distributed disk-related activity rather than
a localized nuclear starburst  \citep[e.g. see][]{irw11}, but the physics may be similar.
 For NGC~253, strong electron energy losses are implied, for example. 

We defer the point-by-point decomposition of the
map into thermal and non-thermal components to a future paper; nevertheless, it is
possible to see
evidence for the dominance of thermal emission in at least two specific
star forming regions, best seen at the highest resolution spectral index maps
(Fig.~\ref{fig:alpha}).
 One has been pointed out earlier (Sect.~\ref{sec:results_total})
and is marked with a cross on the H$\alpha$ map of 
Fig.~\ref{fig:C-band}c as well as the high resolution
spectral index map of Fig.~\ref{fig:alpha}b.  The spectral index at this position
is $\alpha$ = -0.16 $\pm$ 0.09 (averaged from 
Fig.~\ref{fig:alpha}b over the star forming region)
which is entirely consistent with a thermal spectrum, within uncertainties.
Another star forming region located 1 arcmin to the SSW of this location 
and visible as a discrete peak in Fig.~\ref{fig:C-band}c also
shows a similarly flat index, consistent with thermal emission.

\section{Conclusions}
\label{sec:conclusions}

In this paper, we have presented the first results from a new survey, called CHANG-ES,
to observe radio continuum halos in 35 edge-on, normal spiral galaxies. The galaxies
are being observed in all polarization products in two different bands, 1.5 GHz (L-band) and 6 GHz
(C-band), and over 3 different EVLA array configurations.  This is the first comprehensive
radio continuum survey of halos to include all polarization products.  The motivation and
science goals for the survey have been presented in Paper I.
 
Our initial CHANG-ES test observations of NGC~4631 have been carried out at C array alone and therefore 
are sensitive to the disk, the disk-halo interface, and inner halo emission; we do
not detect the faint large scale halo emission because of the lack of
large scale sensitivity in this array configuration.   Our new results demonstrate that, even with
modest integration times, new details of the emission in this galaxy have emerged.  

One advantage of the wide-band, multi-channel EVLA capabilities 
 is the ability to form spectral index maps at a common spatial
resolution {\it within} a single observing band in a single array configuration.  We have
formed such maps at L-band.    Our  
 L-band spectral index map has the highest resolution yet obtained for
this galaxy and is the first in-band spectral index map published for an edge-on galaxy. 
In addition, with appropriate uv weighting, the EVLA wide bands have allowed us to match spatial 
resolution between
L-band and C-band in a single array configuration.

Our results for NGC~4631 include:

$\bullet$ At both 1.5 and 6 GHz, numerous extensions can be seen emerging from the disk into the
halo, many of which have not previously been observed.  

\begin{figure*}[hbt]
\setlength{\unitlength}{0.1in}
\begin{picture}(65,75)
\put(20,50){\includegraphics{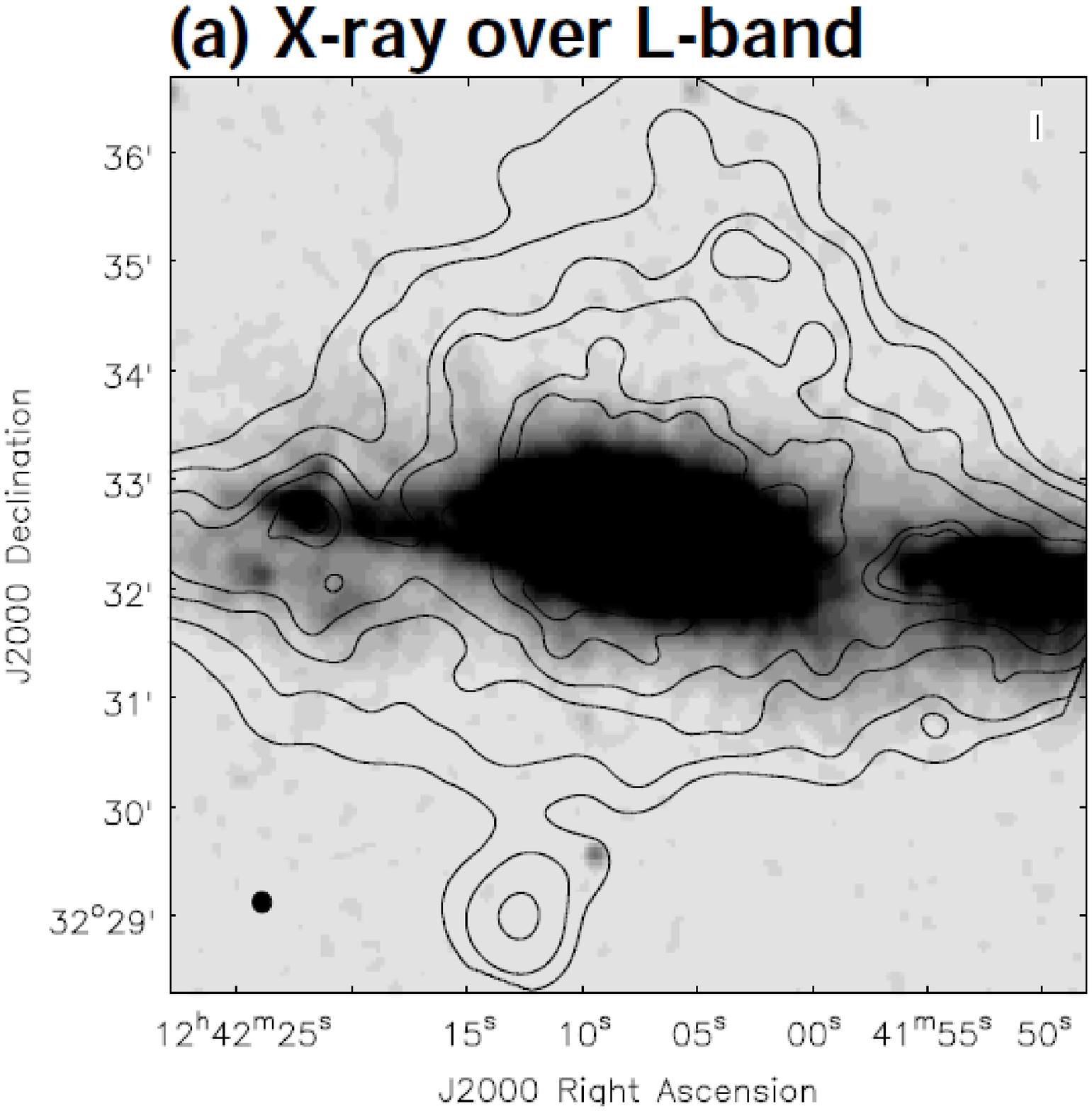}}
\put(16,18){\includegraphics{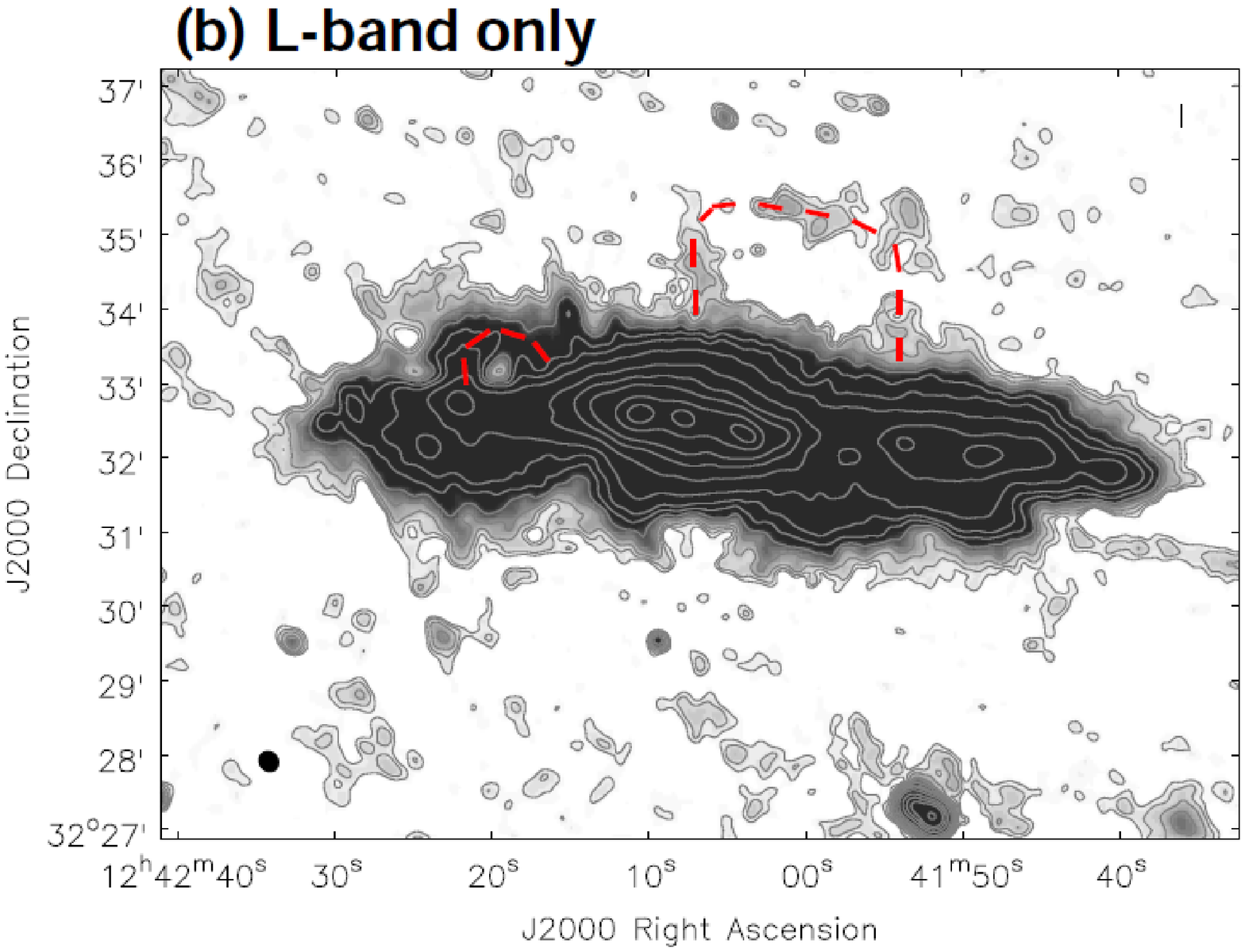}}
\put(15,-13){\includegraphics{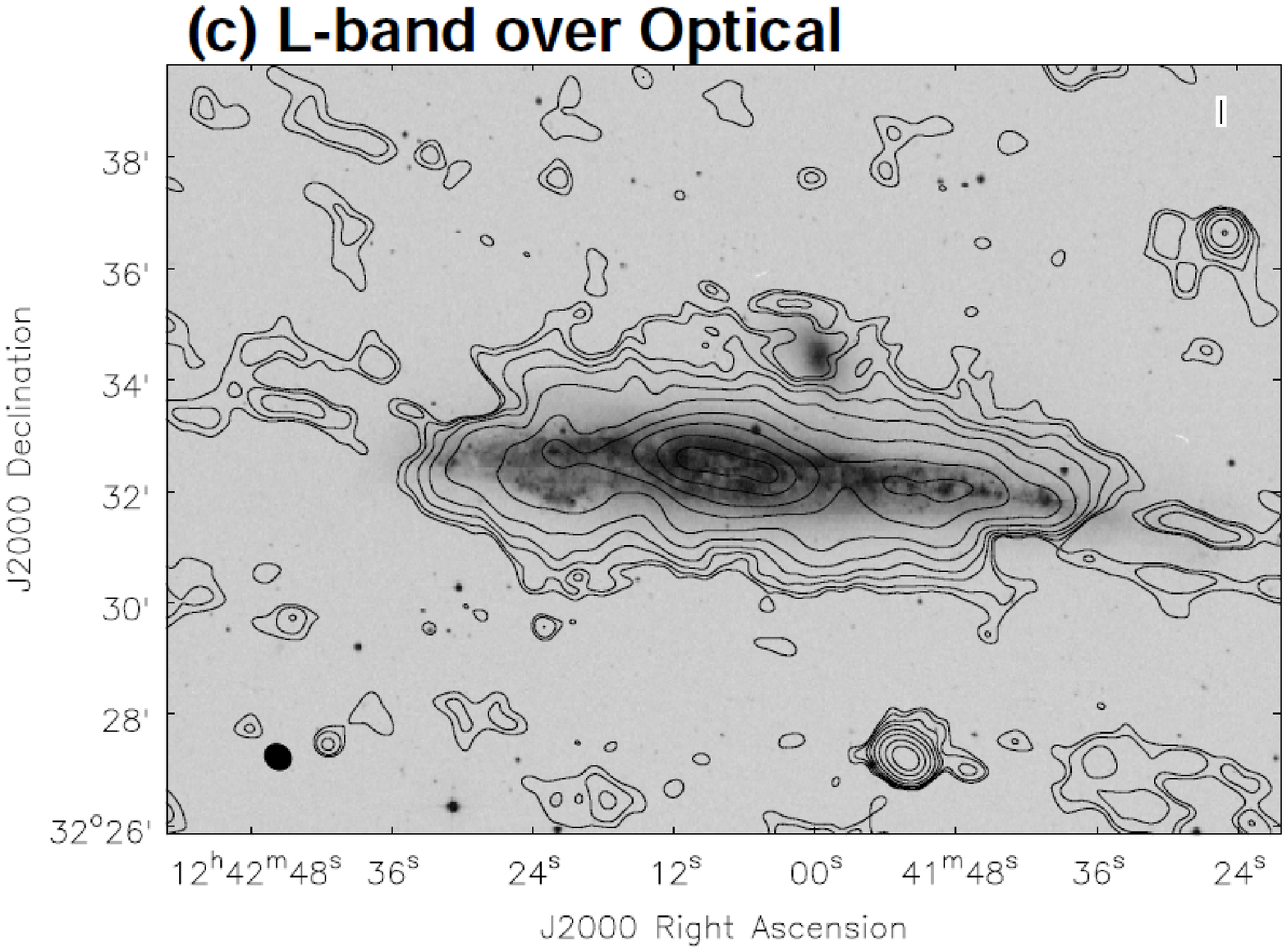}}
\end{picture}
\caption{L-band (1.5 GHz) images of NGC~4631. 
The beam is shown as the filled ellipse at lower left.
 See Table~\ref{table:map-n4631} for map parameters
{\bf (a)}: Contours of soft X-ray emission \citep{wan01} over the Robust 0 weighted
greyscale image.
 {\bf (b)}: Contours of the Robust 2 weighted image over a self-greyscale.
Contours are at 9 (2$\sigma$), 13.5, 20, 30, 45, 65, 90, 150, 300, 500, 750, 1500, and 2500
$\times$ 10 $\mu$Jy beam$^{-1}$. Red dashed curves denote the loops discussed in
Sect.~\ref{sec:results}. 
{\bf (c)}:
 Contours of the uv tapered image over the Second Digitized Sky Survey (DSS2) blue  
image.   Contours are at
2 (2$\sigma$), 3, 5, 8, 15, 30, 65, 120, 250, and 500 $\times$ 100
$\mu$Jy beam$^{-1}$. }
\label{fig:L-band}
\end{figure*}

$\bullet$ At 1.5 GHz, two extra-planar features appear to form
loops in projection.  The larger loop is 6.3 kpc in diameter and located on the north side of
the disk slightly to the west of center.  The smaller loop, 2.6 kpc in diameter,
is also on the north side of the disk
but towards the east (see Fig.~\ref{fig:L-band}b).

$\bullet$ The larger loop is exterior to the soft X-ray feature observed by \citet{wan01}
(Fig.~\ref{fig:L-band_xray}).  The
 minimum energy magnetic field is 8 $\mu$G and the
magnetic pressure in this loop ($P_{mag}\,=\,3\,\times\,10^{-12}$ erg cm$^{-3}$) is sufficient to
constrain the hot X-ray gas ($P_{th}\,=\,6.9\,\times\,10^{-13}$ erg cm$^{-3}$).

$\bullet$ The smaller loop is approximately at the location of the HI worm and supershell identified by
\citet{ran93}.  The spectral index within and adjacent to the loop is much steeper ($\alpha\,\approx\,-2.1$)
than along the loop itself ($\alpha\,\approx\,-1.3$).

$\bullet$ At 1.5 GHz, 
a spur of polarized emission can be seen away from the plane,
consistent with previous observations; this spur is likely associated with the
eastern loop.

$\bullet$ At 6 GHz, the polarization is higher and the average percentage polarization is 7\% over regions
within which the S/N was high enough to be measured, i.e. predominantly in the disk. 
The {\it apparent} magnetic field orientation in the plane of the disk (uncorrected for Faraday rotation) 
appears to be parallel to the disk. 

$\bullet$ There is one position in the disk, corresponding to the HII region complex, CM67,
 at which the apparent field orientation changes sharply (Fig.~\ref{fig:percentpol}); this is a target
region for future in-depth analysis including Faraday rotation.

$\bullet$  At 1.5 GHz, the average spectral index is $\bar\alpha\,=\,-0.84\,\pm\,0.05$, indicating
that the emission is predominantly non-thermal, on average, throughout the disk.  However, a small thermal contribution is 
sufficient to explain the observed positive spectral curvature
in this band.

\begin{figure*}[hbt]
\setlength{\unitlength}{0.1in}
\begin{picture}(65,74)
\put(8,0){\includegraphics{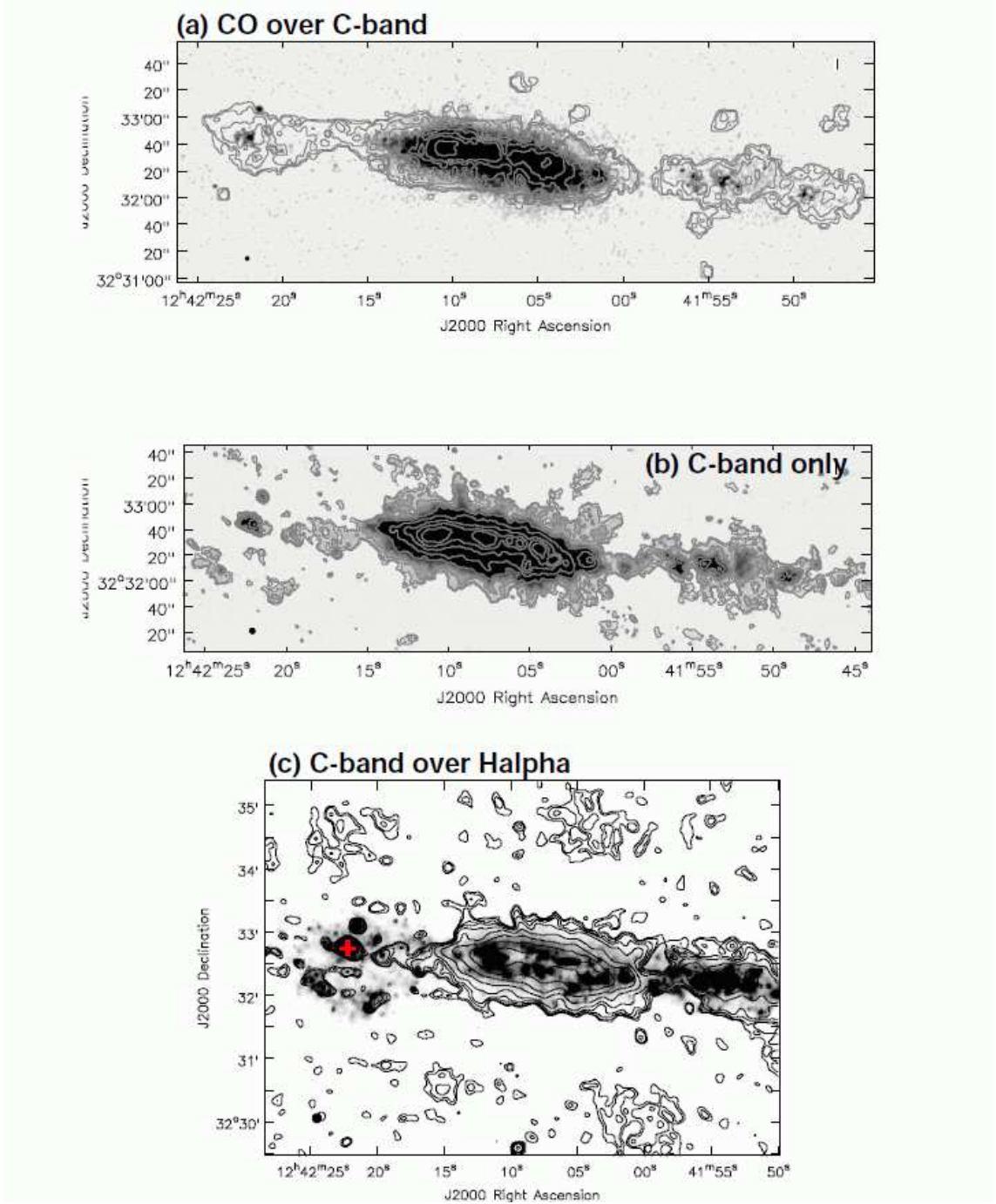}}
\end{picture}
\caption{C-band (6 GHz) images of NGC~4631. 
The beam is shown as the filled ellipse at lower left.
 See Table~\ref{table:map-n4631} for map parameters.
{\bf (a)}: Contours of CO(J=3-2) emission \citep{irw11} over the Robust 0 weighted
image.
{\bf (b)}: Contours of the Robust 2 weighted image over a self-greyscale.
Contours are at
11 (2$\sigma$), 16.5, 30, 50, 80, 150, 300, 500, and 900 $\mu$Jy beam$^{-1}$.
{\bf (c)}: Contours of the uv tapered image over a greyscale
of H$\alpha$ emission from \citet{hoo99}. Contours are at 
16 (2$\sigma$), 24, 40, 60, 100, 200, 400, 800 and 2500 $\mu$Jy beam$^{-1}$.
The cross marks the location of a star formation complex discussed in Sect.~\ref{results:spectral_index}.
}
\label{fig:C-band}
\end{figure*}
\begin{figure*}[hbt]
\setlength{\unitlength}{0.1in}
\begin{picture}(65,74)
\put(2,-23){\includegraphics{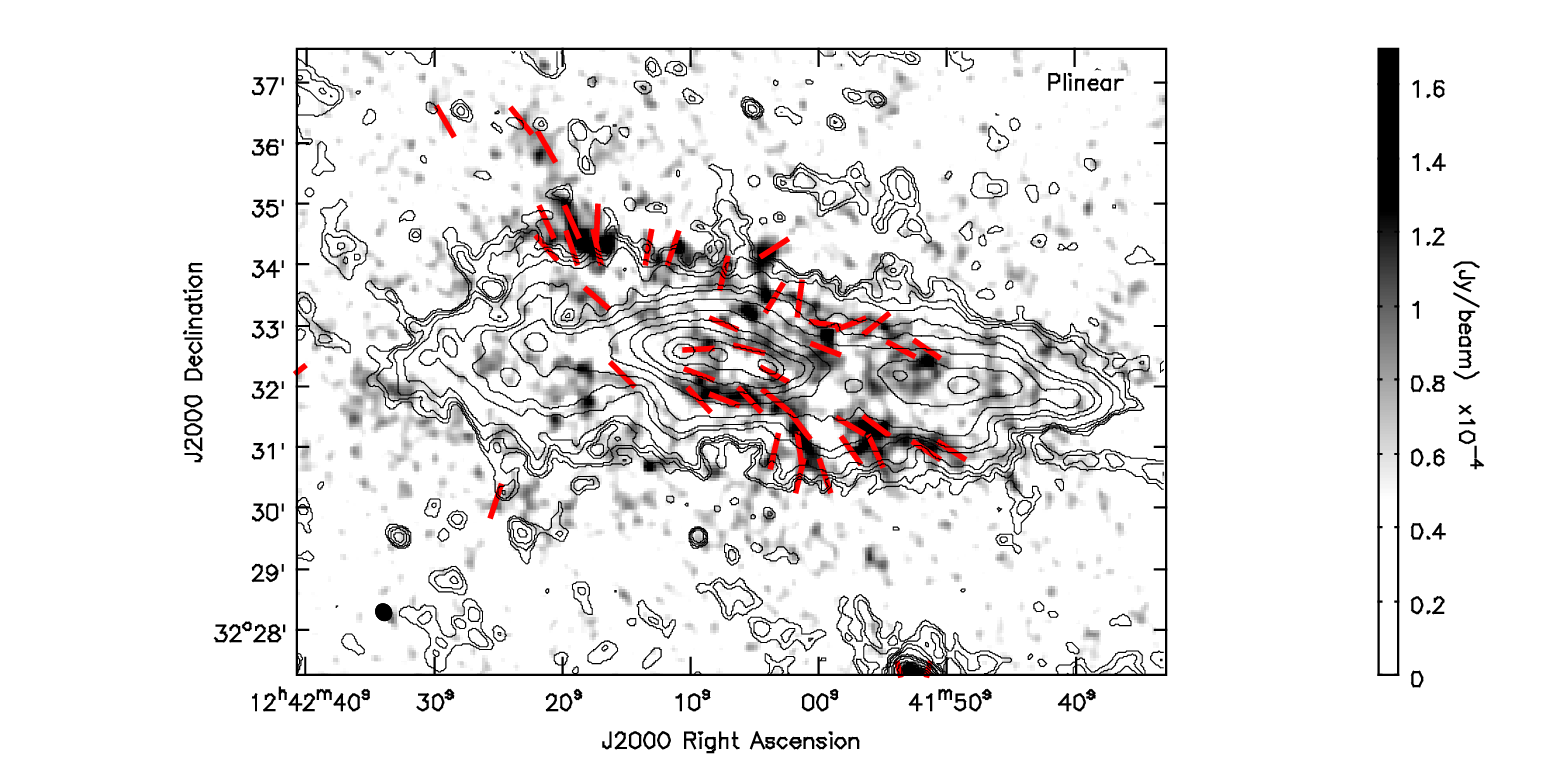}}
\put(-2,-62){\includegraphics{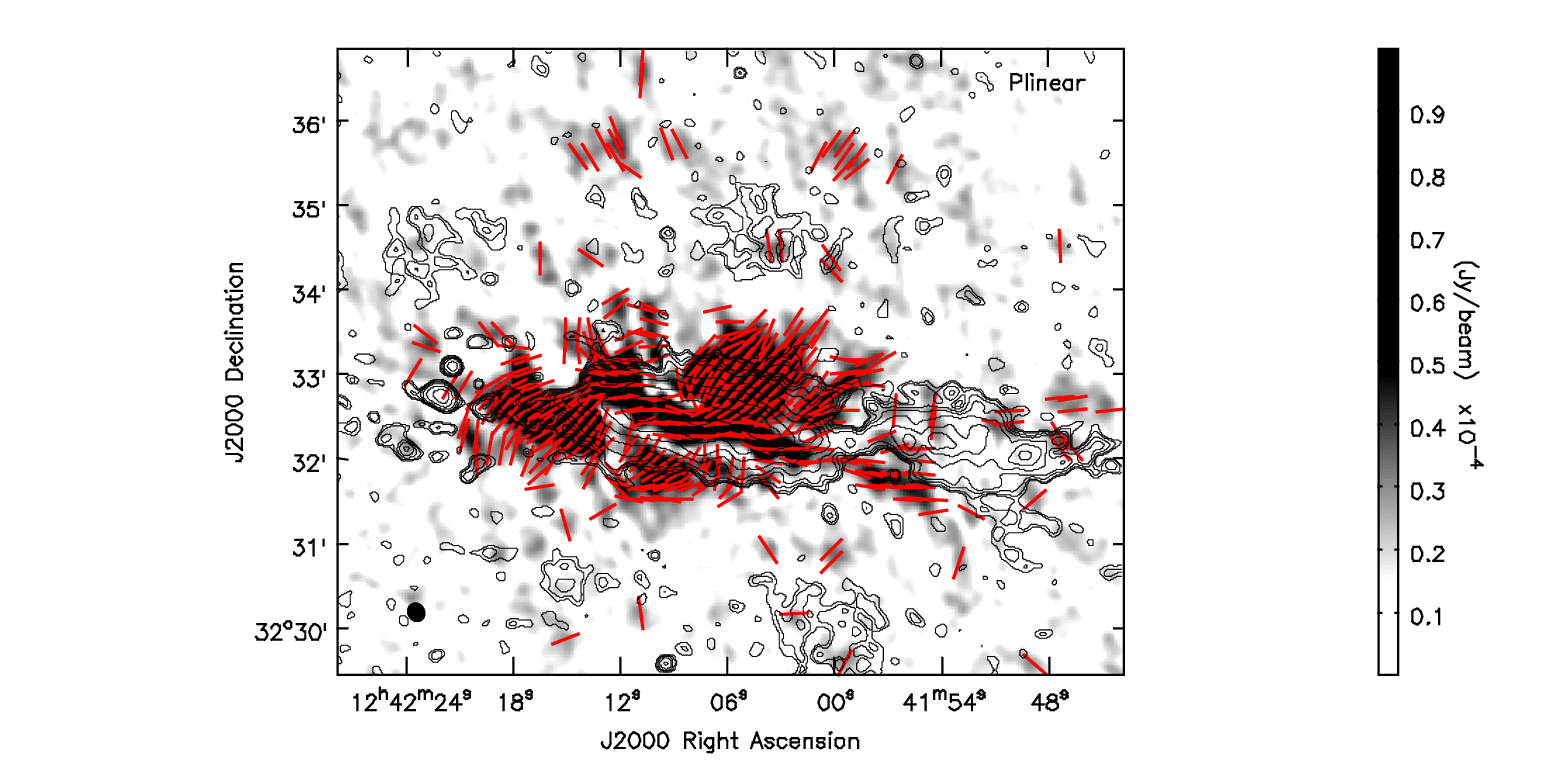}}
\end{picture}
\caption{Polarization maps of NGC~4631.  The linear polarization is shown in greyscale
with the greyscale range shown at right.  Contours show the total intensity emission,
and red vectors show the apparent angle of the magnetic field. 
(Note that the vector length could only be specified as a
 constant in the CASA package.) {\it Maps have been corrected for ionospheric Faraday rotation, but
 not for Faraday rotation intrinsic to the galaxy itself.}
The beam is shown
as a filled ellipse at lower left.
{\bf (a)}  L-band polarization corresponding to Fig.~\ref{fig:L-band}b.
{\bf (b)}  C-band polarization corresponding to Fig.~\ref{fig:C-band}c.
}
\label{fig:pol_maps}
\end{figure*}
\begin{figure*}[hbt]
\setlength{\unitlength}{0.1in}
\begin{picture}(65,74)
\put(8,-55){\includegraphics{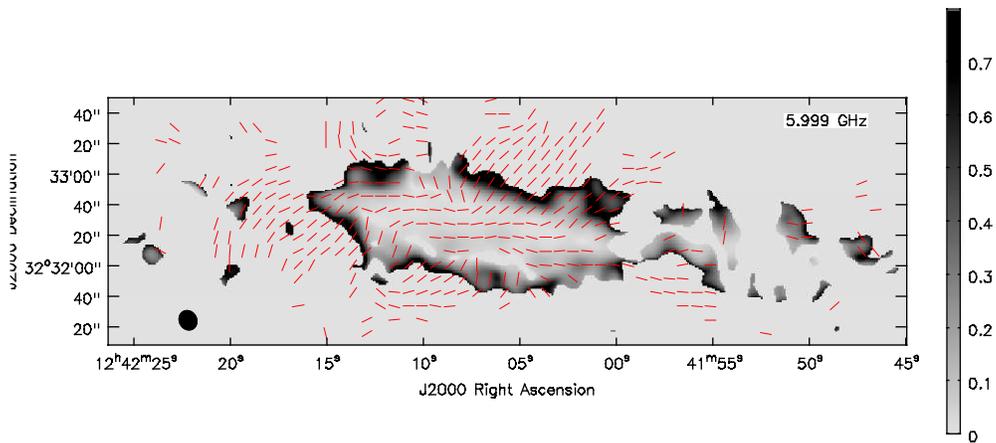}}
\end{picture}
\caption{Map of percentage polarization at C-band superimposed with
the apparent magnetic field vectors as shown in Fig.~\ref{fig:pol_maps}b.  The
greyscale range is shown at right and the beam ellipse is shown at
bottom left.  Notice the sharp change of vector orientation at
RA $\approx$ 12$^{\rm h}$ 42$^{\rm m}$ 12$^{\rm s}$,  DEC $\approx$ 32$^\circ$ 32$^\prime$ 30$^{\prime\prime}$.
}
\label{fig:percentpol}
\end{figure*}
\begin{figure*}[hbt]
\setlength{\unitlength}{0.1in}
\begin{picture}(65,74)
\put(0,-15){\includegraphics{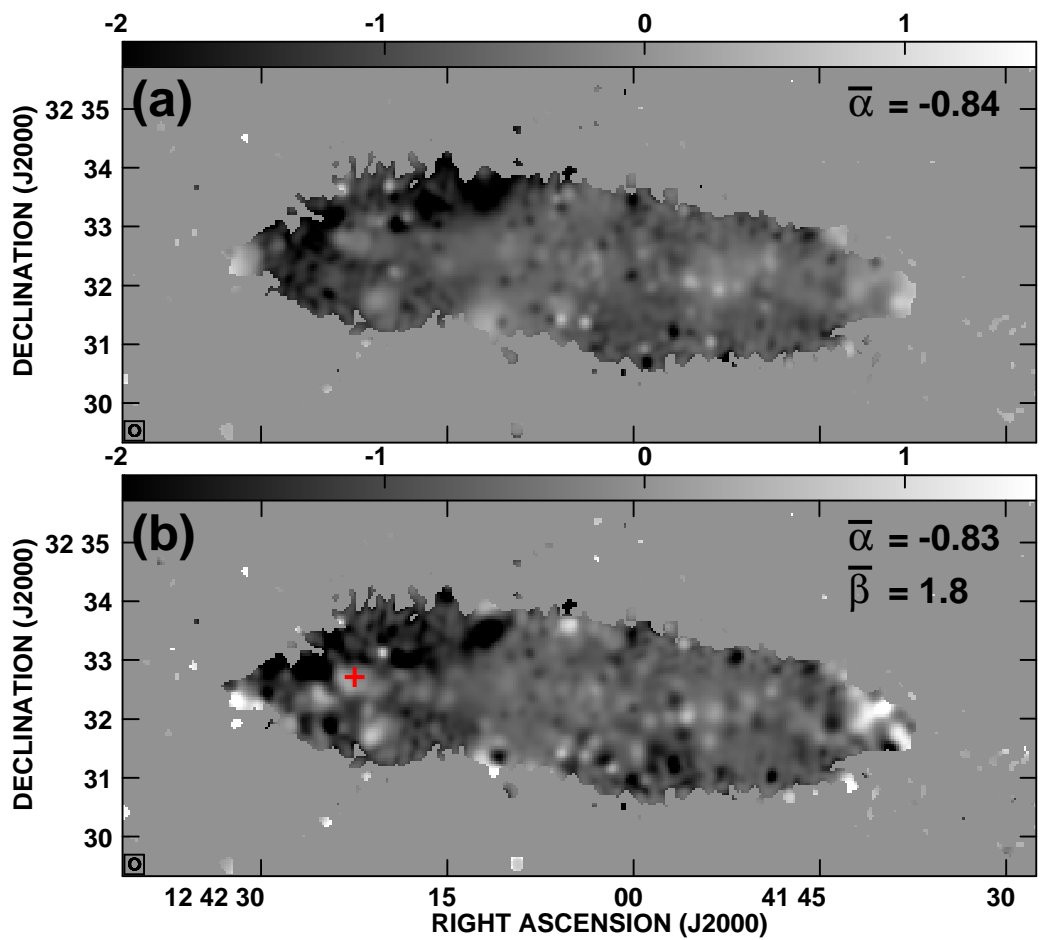}}
\end{picture}
\caption{L-band spectral index maps corresponding to Fig.~\ref{fig:L-band}a,
shown to the same greyscale and spatial scale.
The spectral index and curvature (when fitted), averaged over the
displayed emission, are denoted at upper right. The beam is indicated as an open ellipse
at lower left.  The per-pixel error bar is 0.4 (Sect.~\ref{sec:spectral_index}), creating the
`mottled' appearance; extreme values within a beam width of the perimeter are
artifacts.
{\bf (a)} Fit to Eqn.~\ref{eqn:spectralfit} assuming no curvature ($\beta\,=\,0$).
{\bf (b)} Fit to Eqn.~\ref{eqn:spectralfit}, solving for both $\alpha$ and
$\beta$. The cross marks the location of a star forming complex also marked in
Fig.~\ref{fig:C-band}c.
}
\label{fig:alpha}
\end{figure*}
\begin{figure*}[hbt]
\setlength{\unitlength}{0.1in}
\begin{picture}(65,74)
\put(0,15){\includegraphics{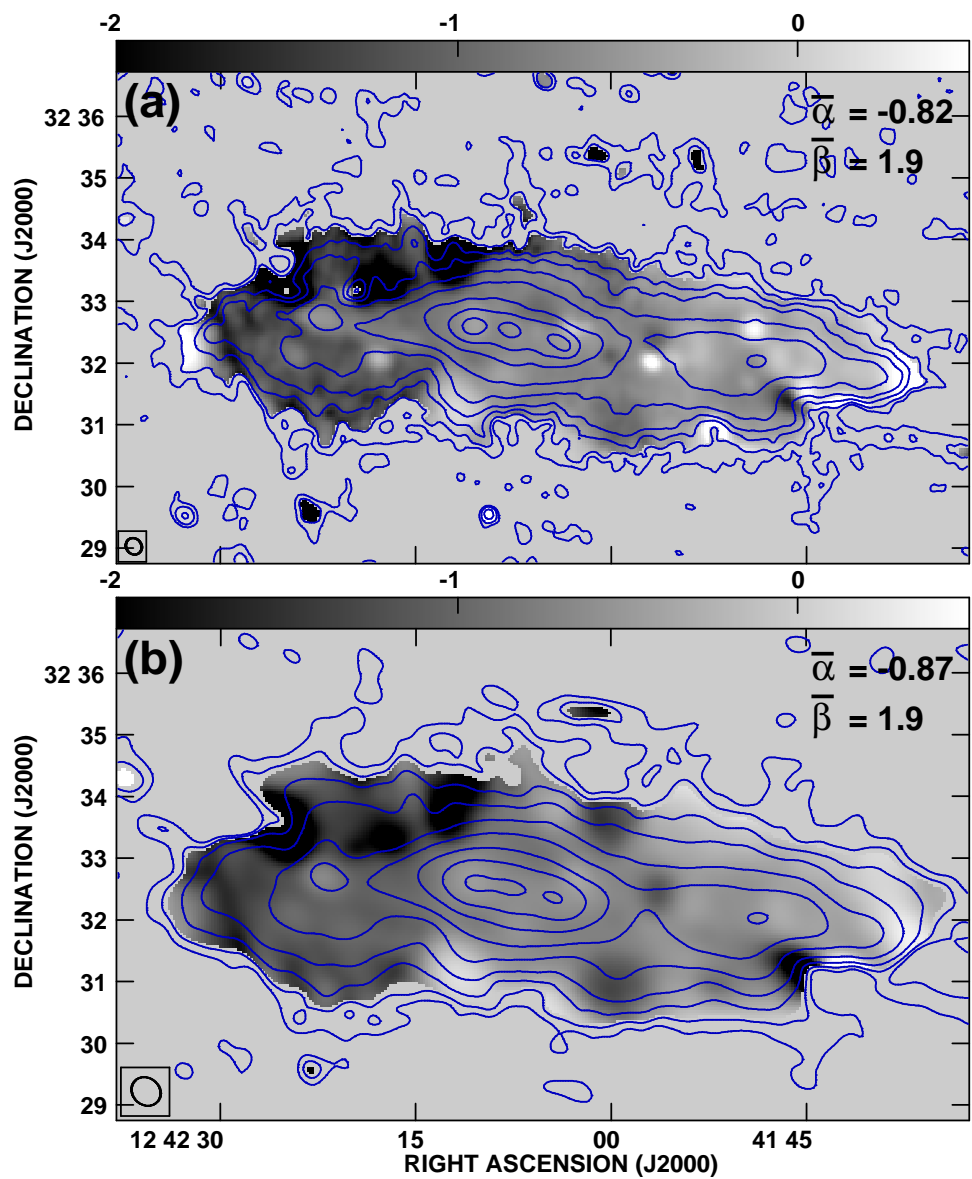}}
\end{picture}
\caption{L-band spectral index maps of Fig.~\ref{fig:L-band}b and c,
shown to the same greyscale (shown at top) and spatial scale.
The spectral index and curvature, averaged over the
displayed emission, are denoted at upper right.  Contours are of the corresponding
total emission.  The beam is indicated as an open ellipse at lower left.
{\bf (a)} Maps corresponding to Fig.~\ref{fig:L-band}b.
Contours are at 9 (2$\sigma$), 20, 40, 65, 125, 250, 600, 1200, and 2400 $\times\,10^{-5}$ Jy beam$^{-1}$. 
{\bf (b)} Maps corresponding to Fig.~\ref{fig:L-band}c.
Contours are at 2 (2$\sigma$), 4, 8, 16, 35, 70, 150, 300 and 600 $\times\,10^{-4}$ Jy beam$^{-1}$.
}
\label{fig:alpha_lowres}
\end{figure*}
\begin{figure*}[hbt]
\setlength{\unitlength}{0.1in}
\begin{picture}(65,74)
\put(-5,-70){\includegraphics{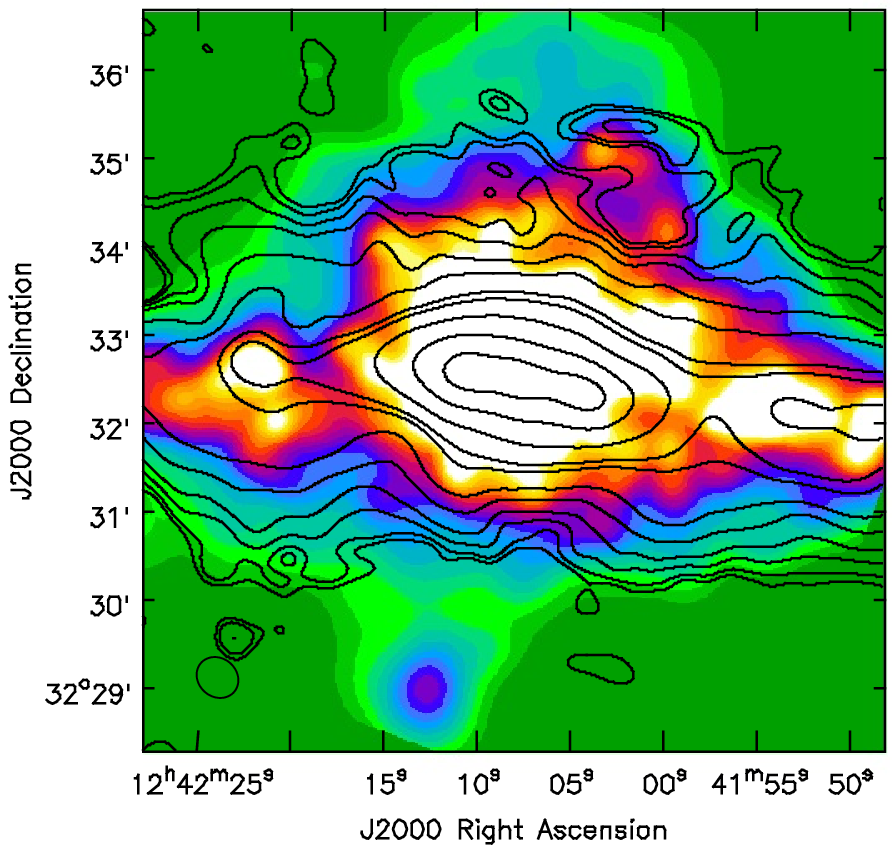}}
\end{picture}
\caption{L-band contours of Fig.~\ref{fig:L-band}c over a colour 
soft X-ray image of Fig.~\ref{fig:L-band}a enhanced to show the
X-ray - radio continuum association in the loop north of the galaxy's
center.
}
\label{fig:L-band_xray}
\end{figure*}

$\bullet$  At specific discrete locations in the disk, it is clear that the thermal contribution is not negligible.
An example is the location of a specific star forming region identified with a cross in Fig.~\ref{fig:C-band}c
whose flat 1.5 GHz spectral index is consistent with thermal emission.

\clearpage

\clearpage

\clearpage

\clearpage

\clearpage

\clearpage

\clearpage

\acknowledgments

JAI and DJS would like to thank the staff at the EVLA for their warm welcome and assistance
during their sojourn in Socorro.  Research at Ruhr-Universit{\"a}t, Bochum, is supported by 
Deutsche Forschungsgemeinschaft
 through grants, FOR1048 and FOR1254.
The Digitized Sky Surveys were produced at the Space Telescope Science Institute under 
U.S. Government grant NAG W-2166. The Second Palomar Observatory Sky Survey (POSS-II) 
was made by the California Institute of Technology with funds from the National Science 
Foundation, the National Geographic Society, the Sloan Foundation, the Samuel 
Oschin Foundation, and the Eastman Kodak Corporation. 
The National Radio Astronomy Observatory is a facility of the National
Science Foundation operated under cooperative agreement by Associated
Universities, Inc.



{\it Facilities:} \facility{EVLA} 

\clearpage




\clearpage

\clearpage

\begin{deluxetable}{lcc}
\tabletypesize{\scriptsize}
\renewcommand{\arraystretch}{0.90}
\tablecaption{NGC~4631 Observing Parameters\label{table:obs-n4631}}
\tablewidth{0pt}
\tablehead{
\colhead{Parameter}                   & 1.5 GHz                   & 6.0 GHz 
}
\startdata
Array                                 & C             & C                     \\
No. antennas\tablenotemark{a}         & 25            & 22                      \\
AC Central Frequency (GHz)\tablenotemark{b}   & 1.375         & 5.5                   \\
BD Central Frequency (GHz)\tablenotemark{b}   & 1.625         & 6.5                    \\
Spw bandwidth (MHz)                   & 32            & 128                    \\
No. spws (AC plus BD)\tablenotemark{c}& 16            & 16                      \\
Total bandwidth (AC plus BD)(GHz)     & 0.512         & 2.048                  \\
No. channels/spw                      & 64            & 64                     \\
Total no. channels                    & 1024          & 1024                    \\
Channel separation (MHz)              & 0.500         & 2.00                  \\
Integration time (s)\tablenotemark{d} & 10            & 10                      \\ 
Obs. Time (min)\tablenotemark{e}      & 30            & 75                      \\
Flux calibrator\tablenotemark{f}       & 3C286        & 3C286                  \\
Zero-polarization calibrator          & OQ208         & OQ208                  \\
Phase calibrator                      & J1221+2813    & J1221+2813             \\
\enddata
\tablenotetext{a}{The total number of antennas for which data were acquired during the
observing session, after problem antennas were flagged.}
\tablenotetext{b}{AC and BD IFs are contiguous in frequency.}
\tablenotetext{c}{Number of spectral windows in AC or BD IFs.}
\tablenotetext{d}{Single record measurement time.}
\tablenotetext{e}{On-source observing time before flagging.}
\tablenotetext{f}{This source was also used as the bandpass calibrator and for determining the
absolute position angle for polarization.}
\end{deluxetable}

\clearpage

\begin{deluxetable}{lcccccc}
\tabletypesize{\scriptsize}
\rotate
\tablecaption{NGC~4631 Map Parameters\label{table:map-n4631}}
\tablewidth{0pt}
\tablehead{
\colhead{Parameter}  & \multicolumn{3}{c}{1.5 GHz}  & \multicolumn{3}{c}{6.0 GHz} 
}
\startdata
uv weighting\tablenotemark{a}        & Briggs Rob = 0  & Briggs Rob = 2 & Briggs Rob = 2 
& Briggs Rob = 0  & Briggs Rob = 2 & Briggs Rob = 0   \\
uv taper (k$\lambda$)\tablenotemark{b}  & none            & none           & 5 
 & none            & none           & 10   \\
Nterms\tablenotemark{c}  &   3            &     3             & 3
 &  3                &  3           &     3            \\
No. Self-cals\tablenotemark{d} &  1 a\&p        &  1 a\&p           & 1 a\&p
 & 1 a\&p            &  1 a\&p           & 1 a\&p      \\
\hline
 & \multicolumn{6}{c}{I images}\\
\hline
Figure label\tablenotemark{e} & Fig.~\ref{fig:L-band}(a) &
 Fig.~\ref{fig:L-band}(b) & Fig.~\ref{fig:L-band}(c) &  Fig.~\ref{fig:C-band}(a) & Fig.~\ref{fig:C-band}(b) & 
 Fig.~\ref{fig:C-band}(c)\\
Synth. beam\tablenotemark{f} \\
~~~~~~($^{\prime\prime}$, $^{\prime\prime}$, $^\circ$)  & 11.16, 10.28, -175.0
& 16.69, 14.77, 34.5
& 29.90, 26.08, 48.9
& 2.71, 2.63, -50.9
& 4.07, 3.80, 27.5
& 8.60, 8.52, 41.5
\\
rms ($\mu$Jy beam$^{-1}$)\tablenotemark{g} & 28 & 45 & 100 & 4.8 & 5.5 & 8.0\\
\hline
 & \multicolumn{6}{c}{Q \& U images\tablenotemark{h}}\\
\hline
Synth. beam\tablenotemark{f} \\
~~~~~~($^{\prime\prime}$, $^{\prime\prime}$, $^\circ$)  &
 11.01, 10.13,  -176.8 & 12.28, 14.41, 33.7 & 29.35,  25.51, at 48.3 &
2.71, 2.63, -50.8 & 4.07 x 3.80 at 27.5 & 13.15, 11.45, 27.2\\
rms ($\mu$Jy beam$^{-1}$)\tablenotemark{g} & 27 & 22 & 23 &
4.7 & 3.5 & 7.0\\
\enddata
\tablecomments{These parameters represent
the images presented in Figs.~\ref{fig:L-band} and \ref{fig:C-band}.}
\tablenotetext{a}{See \citet{bri95} for a description of Briggs weighting with various
`robust' factors.}
\tablenotetext{b}{Scale length of Gaussian taper applied in the uv plane.}
\tablenotetext{c}{Number of terms in the Taylor expansion for fitting the spectral index.}
\tablenotetext{d}{Number of self-calibration iterations, where `a\&p' refers to amplitude and
phase together.}
\tablenotetext{e}{Map labels, as shown in Figs.~\ref{fig:L-band} and \ref{fig:C-band}.}
\tablenotetext{f}{Synthesized beam major and minor axis and position angle.}
\tablenotetext{g}{Rms map noise before primary beam correction.}
\tablenotetext{h}{The cross-hands (RL, LR) had different sets of flags applied than
the parallel hands (RR, LL), leading to differences in the synthesized beams and 
noise for Q and U compared to I.  Cleaning also proceeded with different scales.}
\end{deluxetable}

\clearpage





\end{document}